\documentclass[a4paper,11pt]{article}
\pdfoutput=1 

\usepackage{jheppub} 
\usepackage{soul}
\usepackage{xcolor}

\usepackage{empheq}
\usepackage{mathrsfs} 
\usepackage{url}
\usepackage[utf8]{inputenc}
\usepackage[toc,page]{appendix}
\usepackage[autostyle]{csquotes}
\usepackage{slashed}
\usepackage{amssymb}
\usepackage{graphicx}
\usepackage{longtable}
\usepackage{array}
\usepackage{bm} 
\usepackage{amsmath}
\usepackage{nccmath}
\usepackage{cancel}
\usepackage{mathtools}
\usepackage{amsfonts}
\usepackage{amssymb}
\usepackage[labelfont=bf]{caption}
\usepackage{paralist} 
\usepackage{hyperref}
\usepackage{cleveref}
\DeclareMathAlphabet{\mathpzc}{OT1}{pzc}{m}{it} 
\usepackage{tensor}
\usepackage{stmaryrd}
\title{\boldmath A time-like window into tensionless worldsheets}

\author[]{Sudip Karan}
\author[]{and Bibhas Ranjan Majhi}
\affiliation[]{Department of Physics, Indian Institute of Technology Guwahati,\\ Guwahati 781039, Assam, India.}

\emailAdd{sudip.karan@iitg.ac.in}
\emailAdd{bibhas.majhi@iitg.ac.in}

\abstract{
Rindler worldsheets are well known to transition into a tensionless regime in the infinite acceleration limit, acquiring an emergent Carrollian structure. In this work, we uncover an entirely new manifestation of such tensionless dynamics by constructing a time-evolving Milne worldsheet within the future (expanding) and past (contracting) regions of the background target spacetime. We show that as the Milne worldsheet approaches its null horizons, it reveals a deep and previously unexplored interplay between tensionless string dynamics and an emergent Carrollian structure, governed by an ultra-high frequency limit in time evolution. Our results demonstrate that non-inertial string dynamics can emerge in two structurally independent yet complementary settings—governed either by worldsheet acceleration (Rindler) or by the frequency of its intrinsic time evolution (Milne). Remarkably, we uncover a non-trivial duality-like correspondence between Rindler and Milne worldsheets near their respective null horizons, both exhibiting identical tensionless physics despite their distinct causal and dynamical origins. This discovery positions the Milne construction as a genuine time-like counterpart to the ultra-relativistic (Carrollian) or infinite-boost limit, offering new foundational insights into the deep structure of non-inertial string theory.

}

\begin{document} 
	\maketitle
	\flushbottom
	
	
\section{Introduction}\label{intro}

Carrollian structures and Carrollian physics have emerged as compelling tools with a wide range of applications. For comprehensive details of these developments, we refer to \cite{deBoer:2023fnj,Hansen:2021fxi}. The so-called Carroll regime is fundamentally tied to two complementary approaches. First, it naturally arises in the ultra-relativistic limit, where the speed of light ($c$) approaches zero, $c \to 0$ \cite{Levy-Leblond:1965,SenGupta:1966qer}.\footnote{Readers should note that if the system in question has a characteristic velocity $v_c$, the ultra-relativistic limit for the Carroll regime should be interpreted as $c/v_c \to 0$.}  As a related aspect,  Carroll gravity can be derived via a systematic `small $c$' expansion of general relativity (GR) \cite{Hansen:2021fxi}, which can be viewed as a perturbative expansion around the Carroll point $c \to 0$.\footnote{An equivalent approach to the `small $c$' expansion involves setting $c = \epsilon \hat{c}$, where $\epsilon$ is a dimensionless parameter, and expanding around $\epsilon \to 0$.} Alternatively, Carrollian geometry emerges on null hypersurfaces, such as event horizons or conformal boundaries, due to the degenerate nature of the induced spacetime metric \cite{Duval:2014uva,Duval:2014uoa,Hartong:2015xda}. These insights have found extensive application in string theory, particularly in the study of tensionless limits at high energies \cite{Bagchi:2013bga,Bagchi:2015nca,Bagchi:2020fpr}, string dynamics in black hole spacetimes \cite{Bagchi:2021ban,Bagchi:2023cfp,Bagchi:2024rje}, and microstate modeling of black holes \cite{Bagchi:2022iqb}.

To date, string theory remains arguably the most promising framework for constructing a quantum theory of gravity. It generalizes the concept of point particles (as employed in quantum field theory) to one-dimensional fundamental strings, thereby offering a natural basis for a quantum theory of GR. These strings are characterized by an intrinsic length scale $\ell_s$, which determines their tension through the relation $\alpha^\prime \sim \ell_s^2$ (e.g., see \eqref{MW2}). In non-interacting regime, the tension of strings acts as the sole free parameter and leads to two extreme limits. The infinite tension limit ($\mathcal{T} \to \infty$) corresponds to the point-particle limit ($\ell_s \to 0$), where strings lose their distinctive `stringy' nature ($\alpha^\prime \to 0$), and string theory reduces to the low-energy sector of GR and supergravity. Conversely, the tensionless limit ($\mathcal{T} \to 0$) propels strings into an `ultra-stringy' regime ($\alpha^\prime \to \infty$), where they become infinitely stretchable ($\ell_s \to \infty$), pushing the framework into its ultra-high-energy sector, far from the end of GR and supergravity. Notably, the tensionless limit bears some similarities to the point-particle limit, akin to the massless limit in point-particles, where the string worldsheet becomes null \cite{Schild:1976vq} and corresponds to the ``null'' sector of string theory. In this context, the connection between null strings and Ambitwistor strings \cite{Casali:2016atr,Casali:2017zkz} is also noteworthy. All of this underscores the significance of investigating the tensionless limit of string theory to gain a deeper understanding of how strings perceive generic null surfaces and, possibly the spacetime singularities.

This work focuses on exploring the broad direction of probing the tensionless regime of closed-string worldsheets through reaching their Carrollian point. Carrollian strings naturally arise on the worldsheet when their tension vanishes \cite{Bagchi:2013bga} or when the background target spacetime possesses a Carrollian structure \cite{Cardona:2016ytk}. Detailed studies of this $\text{tensionless}\leftrightarrow\text{Carrollian}$ regime, both classical and quantum, have revealed intriguing aspects of string theory to date.\footnote{For details, readers are referred to the discussions in \cite{Bagchi:2013bga,Bagchi:2015nca,Bagchi:2016yyf,Bagchi:2017cte,Bagchi:2019cay,Banerjee:2023ekd,Banerjee:2024fbi,Bagchi:2020fpr,Cardona:2016ytk} and the references therein.} A key insight is that the degrees of freedom underlying tensionless strings are fundamentally distinct from those in their usual tensile state. Interestingly, the Bogoliubov transformations (involving quantum modes, vacuum states, and oscillators) on tensile worldsheets at the Carrollian limit help probe into their tensionless regime. 

It has long been speculated (e.g., see \cite{Francia:2007qt}) that the distinction between closed and open strings becomes blurred as soon as the tensionless limit is approached. Bagchi et al. in \cite{Bagchi:2015nca,Bagchi:2019cay} provided a concrete demonstration of how an open string description and related symmetries emerge from closed strings in the tensionless limit, which is contrary to the conventional wisdom. They discovered the existence of a Bose-Einstein condensation of all perturbative closed string degrees of freedom onto the emergent open string, further speculating a new phase transition at the point of vanishing string tension \cite{Bagchi:2019cay}. Furthermore, a picture of ``null string complementarity'' has been observed for accelerated Rindler worldsheets approaching their tensionless limit \cite{Bagchi:2020ats,Bagchi:2021ban}. Interestingly, all these manifestations of the closed-to-open string transition are deeply connected to strings near the Hagedorn temperature where a novel phase transition introduces new degrees of freedom \cite{Atick:1988si}. At this extreme high-temperature limit of string theory, strings effectively become tensionless \cite{Pisarski:1982cn,Olesen:1985ej}, favoring the formation of long strings over heating up a gas of strings \cite{Bowick:1989us,Giddings:1989xe}. Recently, investigation of the non-inertial worldsheets at the quantum level has become a center of interest \cite{Bagchi:2024tyq}. However, to date, there has been no independent or intrinsic investigation of tensionless physics on time-evolving worldsheets as a novel non-inertial branch, in contrast to the established framework based on the null or Carrollian limit of Rindler worldsheets studied in \cite{Bagchi:2020ats,Bagchi:2021ban}. Building on these insights, we aim to extend the existing Carrollian analysis of inertial (Minkowski) and accelerated (Rindler) worldsheets to the tensionless dynamics of a time-evolving worldsheet---referred to as the `Milne worldsheet' in this paper. 

More precisely, the aim of this work is motivated by the following critical aspects and fundamental questions that remain unexplored:

\begin{itemize}

\item Let us consider three distinct sets of observers—$O_{\mathbb{M}}$, $O_{\mathbb{R}}$, and $O_{\mathbb{K}}$—residing on Minkowski, Rindler, and Milne worldsheets, respectively. Naturally, $O_{\mathbb{M}}$ has full access to the entire background target spacetime, which is partitioned into four causally disconnected regions: F (future), P (past), R (right), and L (left), as illustrated in \cref{fig:x}. However, from the perspective of $O_{\mathbb{M}}$, the observer $O_{\mathbb{R}}$ follows well-known hyperbolic trajectories and remains strictly confined to either the space-like separated R or L Rindler wedges.  By contrast, $O_{\mathbb{K}}$ is confined to the time-like separated F (top) and P (down) Milne wedges.\footnote{Please bear in mind that the F and P regions are also conventionally referred to as the Future and Past ``Kasner wedges'' in any generic spacetime dimension, as they respectively correspond to expanding and contracting degenerate Kasner universes (e.g., see \cite{Crispino:2007eb,Socolovsky:2013rga,Higuchi:2017gcd,Ueda:2021nln,mukhanov2005physical}, as well as the excellent review \cite{Kumar:2024oxf} and references therein). While all four regions (F, P, R, L) together can be mapped to a global Rindler spacetime via coordinate transformations from Minkowski space, it is conventional to refer only to the static regions R and L as the `Rindler wedges'. However, since the current paper employs a two-dimensional analysis for the string worldsheet formulation, these F and P regions are also commonly referred to as ``Milne wedges'', since the degenerate Kasner geometry in two dimensions exactly reduces to the Milne universe under a suitable change of coordinates (e.g., see \cite{Kofman:2011tr}).} The motion of $O_{\mathbb{K}}$ is inherently non-inertial but follows linear trajectories where time evolves while the space-like coordinate remains fixed. Notably, there exists a special trajectory (e.g., refer to the $\eta = 0$ line in \cref{fig:1}) where $O_{\mathbb{K}}$ remains static in space while still undergoing time evolution—a possibility fundamentally absent in the case of $O_{\mathbb{R}}$. Evidently, neither $O_{\mathbb{R}}$ nor $O_{\mathbb{K}}$ has full access to the entire Minkowski background spacetime. Therefore, from the perspective of the Minkowski worldsheet, non-inertial worldsheets emerge in two distinct yet complementary ways: (i) the Rindler worldsheet, where acceleration generates R-L entanglement between causally disconnected wedges, and (ii) the Milne worldsheet, where time evolution drives F-P entanglement, allowing for both static and non-static configurations. These formulations suggest that non-inertial string dynamics cannot be fully described within a single unified framework but instead depend on the nature of the underlying worldsheet evolution—whether governed by time evolution or acceleration. More technically, it is well recognized that the physics of a target spacetime is inherently imprinted onto its associated worldsheet. In this context, the Minkowski target spacetime of the worldsheet tied to $O_{\mathbb{M}}$ must be precisely divided—half into the R and L regions and half into the F and P regions—between the worldsheets of $O_{\mathbb{R}}$ and $O_{\mathbb{K}}$, thereby giving rise to their respective Rindler and Milne target backgrounds. This naturally suggests that the complete description of non-inertial worldsheets is governed by a fundamental 1-to-2 correspondence between $O_{\mathbb{M}}$ and the pair $\left(O_{\mathbb{R}}, O_{\mathbb{K}}\right)$. In other words, the physics of the Milne branch of non-inertial worldsheets is not merely a complementary extension but an essential, missing piece of a more comprehensive picture. Its independent exploration is therefore crucial to achieving a deeper and more unified understanding of non-inertial string dynamics.  
\begin{figure}[t!]
\centering
\includegraphics[width=.55\textwidth]{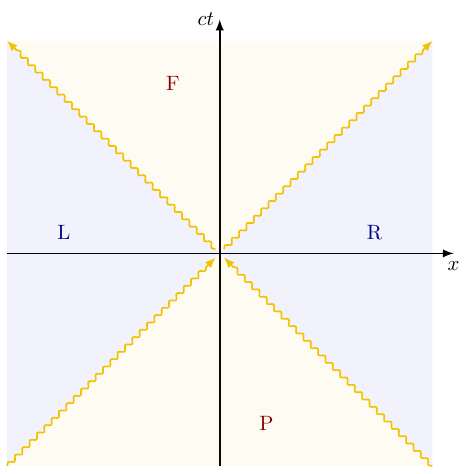}
\caption{A Minkowski spacetime diagram partitioned into four causally disconnected regions: the right (R: $x > c\lvert t \rvert$) and left (L: $x < c\lvert t \rvert$) Rindler wedges, as well as the future (F: $ct > \lvert x \rvert$) and past (P: $ct < \lvert x \rvert$) Milne wedges. Notably, the F and P regions of spacetime correspond to the degenerate Milne universes, exhibiting highly anisotropic expansion and contraction, respectively.}\label{fig:x}
\end{figure}

While the framework of tensile string theory and the tensionless limit on the Rindler worldsheet have been extensively studied in \cite{Bagchi:2020ats,Bagchi:2021ban}, a comprehensive analysis of the Milne worldsheet remains an open problem. The fundamental differences between Rindler and Milne worldsheets introduce an inherent rigidity that prevents a straightforward extrapolation of results from one framework to the other. Moreover, the causal horizons separating Rindler and Milne wedges eliminate any direct communication between $O_{\mathbb{R}}$ and $O_{\mathbb{K}}$ from a field-theoretic perspective, reinforcing the possibility that the physics governing these two worldsheets may be fundamentally different. This naturally leads to the central question: \textit{What governs the dynamics of string theory on a Milne worldsheet? Does it share a deep correspondence with its Rindler counterpart, or do these two frameworks encode entirely distinct physical principles?} Addressing this question requires a non-trivial investigation, as any potential equivalence or disparity between the two worldsheet frameworks must emerge from a deeper analysis of their intrinsic structures, rather than relying solely on their field-theoretic aspects—particularly in terms of quantum entanglement structures, causal properties, and the Bogoliubov transformations governing their quantum modes. The outcome of such an analysis is expected to provide a new perspective on non-inertial string dynamics, offering new insights into the interplay between string theory, quantum field theory, and the causal structure of spacetime.

\item Tensionless worldsheets are known to exhibit a degenerate structure, as exemplified by the Rindler worldsheet when $O_{\mathbb{R}}$ reaches the causal boundary or null horizons from the R or L regions by taking the infinite acceleration limit \cite{Bagchi:2020ats,Bagchi:2021ban}. This tensionless limit is fundamentally linked to an infinite boost, giving rise to an emergent Carrollian structure on the accelerated worldsheet. A distinct yet comparable scenario may occur for $O_{\mathbb{K}}$, where the worldsheet observer approaches the same null horizons from the side of F or P regions, naturally leading to the question of whether a tensionless limit can be realized in the Milne counterparts of non-inertial worldsheets. More precisely, one may ask: \textit{Do the Rindler and Milne worldsheets exhibit identical physics as they approach their respective null horizons? If so, what is the precise nature and interpretation of the limiting process in Milne time evolution that corresponds to the well-known tensionless or Carrollian point attained by taking the infinite acceleration limit in the Rindler case? Can a time-like analog of the Carrollian or infinite boost limit be defined on non-inertial worldsheets?} Investigating these questions requires identifying a characteristic parameter governing time evolution in the Milne worldsheet that plays a role analogous to the acceleration parameter in the Rindler counterpart. A preliminary hint arises from the field-theoretic perspective: the Bogoliubov transformations relating the quantum modes of $O_{\mathbb{R}}$ and $O_{\mathbb{M}}$ are structurally identical to those between $O_{\mathbb{K}}$ and $O_{\mathbb{M}}$, provided an equivalence is established between Rindler acceleration and the frequency of time evolution (e.g., see \cite{Olson:2010jy}). Extending this analysis to string theory will clarify the role of Carrollian structures as probes for tensionless non-inertial worldsheets. If a well-defined time-like Carrollian limit exists, it could transform the existing 1-to-2 correspondence between $O_{\mathbb{M}}$ and the pair $\left(O_{\mathbb{R}}, O_{\mathbb{K}}\right)$ into an emergent 1-to-1 unification. In this scenario, $O_{\mathbb{R}}$ and $O_{\mathbb{K}}$ —which were previously causally disconnected—would gain mutual access to each other’s physics along a shared limiting framework, bridging the gap between acceleration-driven and time-evolution-driven worldsheets. Such a construction would not only solidify the foundational role of the Milne worldsheet but also reveal deeper connections between tensionless string theory, Carrollian limits, and non-Lorentzian extensions of string dynamics.

\item The worldsheet vacuum always plays a pivotal role in testing and verifying string-theoretical predictions and phenomena. The construction of Rindler worldsheets in \cite{Bagchi:2020ats,Bagchi:2021ban} has already demonstrated the emergence of an evolving vacuum that continuously transforms with increasing worldsheet acceleration, allowing the Minkowski vacuum to be expressed as a squeezed or coherent state. Through the lens of quantum field theory, this coherent Minkowski vacuum is often interpreted as an entangled state between the modes residing in the R and L Rindler wedges, and $O_{\mathbb{R}}$ necessarily perceives this acceleration-induced Minkowski vacuum as a thermalized state, in accordance with the Unruh effect \cite{Unruh:1976db}. Interestingly, recent studies have shown that the Minkowski vacuum can also be structured through entangled states spanning the F and P wedges, leading to an alternative realization of the Unruh effect (e.g., see \cite{Olson:2010jy,Higuchi:2017gcd,Ueda:2021nln}). In fact, it has been argued that the typical intrinsic nature of the frequency of the time-evolution (as compared to the Rindler acceleration parameter) enables the timelike Unruh temperature \eqref{UE5} more feasible for detection \cite{Olson:2010jy}. This naturally raises several fundamental questions: \textit{Can the Minkowski vacuum be equivalently constructed on the Milne worldsheet, just as it is on the Rindler worldsheet? Is it possible to achieve a global representation of the Minkowski vacuum by entangling both the Milne and Rindler components? What is the precise interpretation and implication of the notion of time-like Unruh temperature in the context of a non-inertial worldsheet vacuum?} 

Furthermore, both the closed Rindler worldsheet vacua and the induced Minkowski vacuum are known to manifest as open string-bound states of opposite nature at the tensionless limit \cite{Bagchi:2020ats,Bagchi:2021ban}, a phenomenon deeply tied to Hagedorn physics at the extreme high-energy regime of string theory \cite{Atick:1988si,Pisarski:1982cn,Olesen:1985ej}. This naturally prompts a crucial investigation: \textit{Does the same ``null string complementarity'' picture hold between the Milne worldsheet vacua and the corresponding time-like induced Minkowski vacuum at the Carrollian/tensionless limit? Additionally, can the time-like Unruh temperature of non-inertial worldsheet vacua serve as a probe or even a mechanism for triggering the Hagedorn transition in string theory?}

\end{itemize}

Motivated by this,\footnote{This motivation extends beyond the central questions highlighted as well as answered in this paper. Further open problems and promising future directions are outlined in \cref{outlook}.} we now present the technical objective of this paper by briefly elucidating the structure of the tensionless Milne worldsheet in comparison to its Rindler counterpart. The Rindler worldsheet is analogous to an accelerated observer moving within a background target spacetime spanning space-like separated wedges (e.g., R and L in \cref{fig:x,fig:1}). It has been established that, in the limit of infinite acceleration, the Rindler worldsheet manifests to a tensionless state \cite{Bagchi:2020ats,Bagchi:2021ban}. This transition effectively corresponds to reaching the accelerated worldsheet horizon by inducing a degenerate metric structure. Consequently, this links the Carroll limit of the background target spacetime to a Carroll limit of the accelerated worldsheet. In the present work, our technical objectives are threefold. First, we extend the analysis by replacing the target spacetime with one that spans time-like separated wedges (e.g., F and P in \cref{fig:x,fig:1}). This worldsheet setup defines a structure analogous to a time-evolving observer, which we recognize as the \textit{Milne worldsheet}. Next, we embed the standard Milne physics of quantum field theory \cite{Socolovsky:2013rga,Olson:2010jy,Higuchi:2017gcd,Quach:2021vzo,Crispino:2007eb,Ueda:2021nln} into the string worldsheet theory. Specifically, we consider quantum modes living in the two causally disconnected Milne worldsheet wedges. These two sets of modes are inherently independent, yet their entanglement must suffice to describe the entire Milne worldsheet via a quantum mode expansion. Notably, this `time-like entanglement' has been shown to have an interpretation exactly analogous to the conventional entanglement between modes in the Rindler worldsheet wedges \cite{Olson:2010jy} (also see \cite{Quach:2021vzo,Barman:2024dql}). Finally, we hypothesize that the mode functions of the Milne worldsheet must relate to their inertial (i.e., Minkowski) counterparts via Bogoliubov transformations, analogous to the Rindler case, but parameterized by the dynamics of time evolution. Our ultimate goal is to determine whether limiting the characteristic time-evolution parameter---and correspondingly the mode functions and Bogoliubov coefficients--- can enable the construction of a Carrollian structure for the Milne worldsheet. Such a Carrollian structure would be expected to manifest as the `tensionless limit' of time-evolving string worldsheets. In this process, we aim to address all the fundamental questions outlined earlier in our motivation. 

Some aspects of the technical setup and analysis in this paper may appear similar to those in \cite{Bagchi:2020ats,Bagchi:2021ban}, which constructed the tensionless Rindler worldsheet. Admittedly, both studies primarily employ the well-established methodology of Unruh \cite{Unruh:1976db} to derive the necessary Bogoliubov transformations relating the modes and oscillation operators of the non-inertial worldsheet to their Minkowski counterparts. However, the present work extends this framework by providing a more rigorous and generalized treatment of the underlying computations.\footnote{For interested readers, the detailed derivations presented in \cref{KME,KBT,FGM} serve as a systematic introduction to Unruh’s treatment \cite{Unruh:1976db} (also see \cite{book:Birrell}) in the context of string worldsheets. Unlike \cite{Bagchi:2020ats,Bagchi:2021ban}, our analysis does not impose any prior restrictions or set the velocity of light to unity ($c=1$) for Carrollian considerations. More specifically, readers are encouraged to explore this formulation in terms of the standard and elegant light-cone coordinate representation, which allows comprehensive handling of all left- and right-moving worldsheet modes, along with their Hermitian conjugates, across both causally disconnected wedges of the non-inertial worldsheet framework.} Moreover, to facilitate a comparative analysis, the structural organization of the Milne worldsheet framework has been deliberately aligned with that of the well-established Rindler worldsheet \cite{Bagchi:2020ats,Bagchi:2021ban}. Nevertheless, as the paper progresses, we systematically highlight the fundamental differences in interpreting the Milne worldsheet parameters, emphasizing how these distinctions shape its characteristic features and outcome of the current study.

The rest of this paper is organized as follows. In \cref{prelim}, we provide a concise review of the foundational aspects of ordinary Milne physics, a textbook overview of the worldsheet theory of closed tensile strings, and the setup and limits of tensionless string worldsheets. In \cref{wkb}, we introduce the novel framework of time-evolving Milne worldsheets, formulating their quantum mode expansion, quantum vacuum state, and associated Bogoliubov transformations. Subsequently, \cref{TAW} explores the possible limits on a closed Milne worldsheet, leading to the emergence of Carrollian structures and the corresponding tensionless physics. Finally, \cref{diss} summarizes the key findings and discusses the broader implications of the tensionless Milne worldsheets. Given the technical intricacies of the subject, we include \cref{FGM} to present detailed formulations of the `global modes' essential for structuring the Milne worldsheet expansion discussed in \cref{wkb}.

\section{Preliminaries: a quick tour to the basics}\label{prelim}
	
\subsection{The Milne physics}\label{kasner physics}
\begin{figure}[t!]
\centering
\includegraphics[width=.60\textwidth]{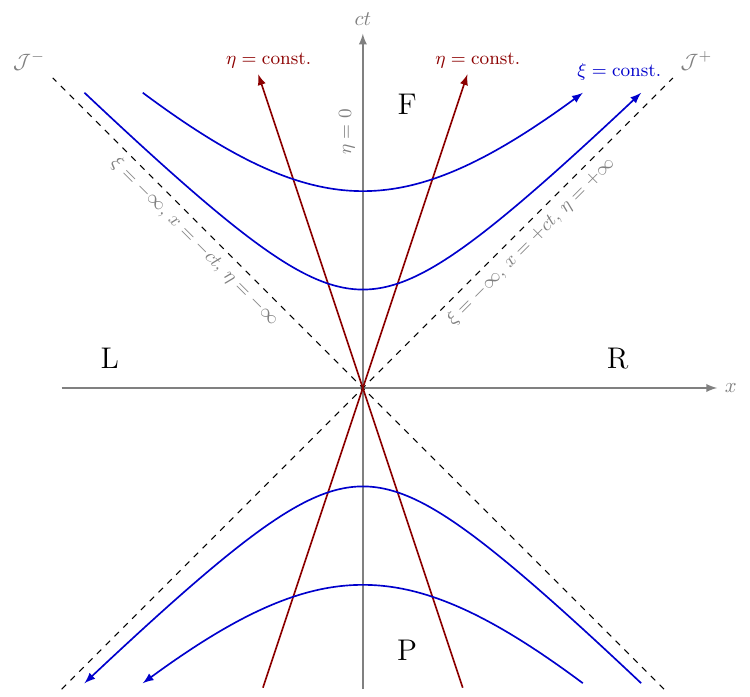}
\caption{The possible trajectories of Minkowski observers in the future (F) and past (P) Milne wedges, with coordinates $(\xi, \eta) \in (-\infty, +\infty)$, bounded by the two null horizons $\mathcal{J}^+ (\xi = -\infty, x = +ct, \eta = +\infty)$ and $\mathcal{J}^- (\xi = -\infty, x = -ct, \eta = -\infty)$. The constant $\eta$ straight lines (in red) represent the only physically allowed trajectories, where the Milne time evolves. Notably, all observers along the $\eta = 0$ line remain static yet non-inertial.}\label{fig:1}
\end{figure} 
The inertial metric describing a Minkowski spacetime in global coordinates is given by
\begin{align}\label{I1}
		ds^2_{\rm M} = c^2dt^2 - dx^2 - d\bm x_\perp^2 ,
\end{align}
where $\bm{x}_\perp$ represents coordinates that are transverse to the plane $(ct, x)$. The top Milne wedge F $(ct > \lvert x \rvert)$ and down Milne wedge P $(ct < \lvert x \rvert)$ as depicted in \cref{fig:1} can be connected to the Minkowski spacetime \eqref{I1} via adopting the following transformations \cite{Socolovsky:2013rga,Olson:2010jy,Higuchi:2017gcd,Quach:2021vzo}
\begin{align}\label{I2}
		\begin{split}
			&{\rm (F) ~:~} 	t =  a^{-1}e^{a\xi}\cosh\left({a\eta/ c}\right),\quad
			x = {ca^{-1}} e^{a\xi} \sinh\left({a\eta/ c}\right),\\[6pt]
			&{\rm (P) ~:~} 	t = -a^{-1} e^{a\xi}\cosh\left({a\eta/ c}\right),\quad
			x = -{ca^{-1}} e^{a\xi} \sinh\left({a\eta/ c}\right).
		\end{split}
\end{align}
Any specific Minkowski observer, satisfying the transformations \eqref{I2} and their inverse \eqref{I4}, moves along a fixed constant $\eta$ coordinate line, intersecting with all the constant $\xi$ hyperbolas in the Milne wedges F and P, as described by:
\begin{align}\label{I5}
	c^2t^2 - x^2  = {c^4\kappa^{-2}}, \qquad {ct}=x\,{\coth}\left({a\eta/ c}\right),
\end{align}
where $a$ is introduced as a real parameter with units of $\rm sec^{-1}$, following the convention in \cite{Olson:2010jy,Quach:2021vzo}. Throughout this paper, we refer to $a$ as the \textit{Milne parameter}. Notably, $\kappa = ac\,e^{-a\xi}$ represents an evolution in time analogous to the proper acceleration for Rindler observers. The coordinates $\xi$ and $\eta$ are adopted respectively as the `time' and `space' coordinates of the Milne spacetime (in both the F and P wedges), resulting in the following metric
\begin{align}\label{I3}
		ds^2_{\rm K} = e^{2a\xi}\left(c^2d\xi^2 - d\eta^2\right) - d\bm x_\perp^2.
\end{align}
The inverse of the transformations \eqref{I2} is given by,
\begin{align}\label{I4}
		\eta = {c\over 2a}\ln \left({ct + x \over ct-x}\right), \quad \xi =  {1\over 2a}\ln \left({a^2 \over c^2}\left(c^2t^2 - x^2 \right)\right).
\end{align}
Therefore, at any moment, one can interpret $ac$ as the redefined time-evolution parameter (analogous to the redefined acceleration for Rindler observers) of an observer in Milne spacetime \eqref{I3} (expanding in F and contracting in P). The Milne- or time-evolution becomes infinite when one hits the null horizons at $\xi \to -\infty$ and $\eta \to \pm \infty$. Notably, for a Rindler observer with a uniform acceleration $a^\prime$ in the R and L wedges (see \cref{fig:x,fig:1}), the scaling parameter $a$ should be substituted by $a^\prime /c$.

Note that here the $\eta=$ constant straight lines are timelike, and therefore, any motion of an object must follow these trajectories. Use of (\ref{I5}) yields,
\begin{equation}
t=\frac{e^{a\xi}}{a}\cosh\left(a\eta/c\right)~.
\label{B1}    
\end{equation} 
So these trajectories can be described by the above relation, where on the right-hand side, only $\xi$ is changing along the path of a moving object. Moreover, in the F sector, as the object moves along these trajectories, the space increases. Therefore, the trajectories can be related to the expansion of space. The P sector corresponds to the reverse one, i.e., contraction of space. 
Moreover, the black hole interior geometry in Kruskal coordinates bears considerable similarity with F and P sectors, except the presence of singularity. Therefore, understanding physics in the F-P sectors illuminate both the Milne and black hole sides.

\subsection{Worldsheet theory of tensile closed strings}\label{wmb}
We start by revisiting a closed tensile bosonic string moving in a $D$-dimensional Minkowski background spacetime with metric $\eta_{\mu\nu}$, ($\mu,\nu = 0,\ldots,D-1$). The related Polyakov action, which describes $D$ massless scalar fields $X^\mu(\sigma^a)$ coupled to the two-dimensional dynamical worldsheet metric $\gamma_{ab}(\sigma^a)$ with coordinates $\sigma^a = (\tau,\sigma)$ for $ a,b = 0, 1$, is given by \cite{Polchinski:1998rq}
\begin{align}\label{MW1}
\tensor*{\mathcal{S}}{_{\rm P}} = -{\frac{\mathcal{T}}{2}}\int d\tau d\sigma \sqrt{-\gamma}\, \gamma^{ab}\partial_aX^{\mu}\partial_bX^{\nu}\eta_{\mu\nu},
\end{align} 
where $\gamma = \det\gamma_{ab}$ and $\mathcal{T}$ is the string tension related to the conventional Regge slope parameter $\alpha^\prime$ (in units of string length squared) as
\begin{align}\label{MW2}
	\mathcal{T} =  {1 \over 2\pi\alpha^\prime}.
\end{align}
In the conformal gauge, where $\gamma_{ab} = \text{diag}(-1,1)$, and under the periodic boundary condition for the closed string (along the spatial coordinate $\sigma$ with arbitrary periodicity $\ell$),
\begin{align}\label{MWx1}
	X^\mu (\tau, \sigma) = X^\mu (\tau, \sigma + \ell),	
\end{align} 
the most general solution to the equations of motion corresponding to \eqref{MW1} is usually expressed in a standard mode expansion, as described in \cite{Polchinski:1998rq}. However, the objective of this work requires restructuring the conventional mode expansion of the worldsheet solution and expressing it in the following form, which resembles a simple Klein-Gordon massless scalar field solution:
\begin{align}\label{MW9}
	X^\mu (\tau, \sigma) = x^\mu + \alpha^\prime p^\mu\tau + \sqrt{2\pi\alpha^\prime}\sum_{n > 0} \left[ \tilde{\alpha}_n^\mu \tilde{\mathcal{V}}_n + \alpha_n^\mu \mathcal{V}_n + \tilde{\alpha}_{-n}^\mu \tilde{\mathcal{V}}_n^* + \alpha_{-n}^\mu \mathcal{V}_n^* \right],
\end{align}
where $\tilde{\alpha}_0^\mu = {\alpha}_0^\mu = \sqrt{\alpha^\prime \over 2}p^\mu$. This solution describes the motion of the string's center-of-mass, with position $x^\mu$ and momentum $p^\mu$, while carrying a superposition of an infinite collection of right- and left-moving oscillation modes, with coefficients $\tilde{\alpha}_n^\mu$ and ${\alpha}_n^\mu$, respectively. The right-moving $\tilde{\mathcal{V}}_n$ and left-moving $\mathcal{V}_n$ modes, along with their Hermitian conjugates $\lbrace \tilde{\mathcal{V}}_n^*, \mathcal{V}_n^*\rbrace$, are successively expressed as
\begin{align}\label{MW10}
	\begin{split}
		\tilde{\mathcal{V}}_n &= {i e^{-i\omega_n (\tau - \sigma/c)} \over \sqrt{4\pi}n}, \qquad \tilde{\mathcal{V}}_n^* = -{i e^{i\omega_n (\tau - \sigma/c)} \over \sqrt{4\pi}n}, \\[10pt]
		\mathcal{V}_n &= {i e^{-i\omega_n (\tau + \sigma/c)} \over \sqrt{4\pi}n}, \qquad \mathcal{V}_n^* = -{i e^{i\omega_n (\tau + \sigma/c)} \over \sqrt{4\pi}n}.
	\end{split}
\end{align}
It is important to note that the parameter $\omega_n = {2\pi cn \over \ell}$ serves as the frequency for the above mode functions on a closed inertial worldsheet with arbitrary periodicity $\ell$, satisfying the closed string condition \eqref{MWx1}. The oscillators $\left\lbrace \tilde{\alpha}_n^\mu, {\alpha}_n^\mu\right\rbrace$ are essentially a rescaled creation operators (for $n <0$) and annihilation operators (for $n >0$), satisfying the commutation relation\footnote{For each mode $n$ and oscillation direction $\mu$, the commutation \eqref{MW7} satisfy a quantum harmonic oscillator algebra with the standard normalization $\tilde{\alpha}_n^\mu \to \sqrt{n}\tilde{\alpha}_n^\mu$, $\tilde{\alpha}_{-n}^\mu \to \sqrt{n}(\tilde{\alpha}_n^\mu)^\dagger$, and similarly for the holomorphic counterparts.}
\begin{align}\label{MW7}
	\left[\tilde{\alpha}_n^\mu, \tilde{\alpha}_m^\nu\right] =  \left[{\alpha}_n^\mu, {\alpha}_m^\nu\right] = n\eta^{\mu\nu}\delta_{n+m,0},
\end{align}
and related via their respective hermitian conjugates $\tilde{\alpha}_{-n}^\mu=(\tilde{\alpha}_n^\mu)^\dagger$ and ${\alpha}_{-n}^\mu=({\alpha}_n^\mu)^\dagger$. Furthermore, they annihilate a Minkowski worldsheet vacuum state $\lvert0_{\rm M}\rangle$, defined to obey:
\begin{align}\label{MW8}
	\tilde{\alpha}_n^\mu\lvert0_{\rm M}\rangle = {\alpha}_n^\mu\lvert0_{\rm M}\rangle = 0, \enspace \forall~ n>0.
\end{align}

\subsection{Tensionless strings and their worldsheet theory}\label{TMW}

In this section, we discuss two widely adopted and well-established approaches for formulating tensionless string theory from its tensile counterpart. The first approach, as outlined in \cite{Isberg:1993av}, involves constructing the action and formulating the tensionless theory from first principles, providing an \textit{intrinsic look at tensionless strings}. The second approach, known as the limiting approach \cite{Bagchi:2013bga,Bagchi:2015nca}, involves taking the appropriate limit on worldsheet coordinates derived from the tensile closed string theory. This limiting process, referred to as \textit{worldsheet contraction}, results in the so-called ``ultra-relativistic limit'' or ``Carrollian limit'' \cite{Duval:2014lpa}, where the worldsheet speed of light tends to zero. Our goal is to briefly revisit both approaches to access their mutual integrity, which will essentially lead us to re-derive the Bogoliubov transformations between the tensionless operators and their tensile counterparts within the context of the inertial Minkowski worldsheet of closed strings.


Revisiting the theory of a tensile worldsheet propagating in a Minkowski target space, one can observe that the action \eqref{MW1} becomes ill-defined when tension $\mathcal{T}$ of fundamental strings vanishes. However, as demonstrated in \cite{Isberg:1993av}, there is a systemic treatment to construct an intrinsic action describing the tensionless string worldsheets,
\begin{align}\label{IL9}
	\mathcal{S}_0 = \int d\tau d\sigma\, V^aV^b h_{ab}, \quad h_{ab}=\partial_aX^{\mu}\partial_bX^{\nu}\eta_{\mu\nu},
\end{align}
where $h_{ab}$ acts as a induced metric which is degenerate or ``null'' (i.e., $\det h_{ab} = 0$) on the tensionless worldsheet. $V^a$ is introduced as a vector density replacing the tensile metric density $\mathcal{T}\sqrt{-\det\gamma_{ab}}\,\gamma^{ab} \to V^aV^b$ at $\mathcal{T}=0$. In the transverse gauge, $V^a = (1,0)$, and by incorporating the closed string periodicity condition \eqref{MWx1}, the field equations describing the tensionless worldsheet \eqref{IL9} were originally solved through a mode expansion involving intrinsic yet anharmonic oscillators \cite{Bagchi:2015nca}. However, it is possible to transform the entire setup into the framework of the standard harmonic oscillator basis, expressing the mode expansion for the tensionless closed Minkowski worldsheet as
\begin{align}\label{IL19}
	X^\mu (\tau, \sigma) = x^\mu + \hat{\alpha}^\prime \hat{p}^\mu\tau + \sqrt{2\pi\hat{\alpha}^\prime}\sum_{n > 0} \left[ \tilde{c}_n^\mu \tilde{V}_n + c_n^\mu V_n + c_{-n}^\mu \tilde{V}_n^* + c_{-n}^\mu V_n^* \right],
\end{align}
where $\tilde{c}_0^\mu = {c}_0^\mu = \sqrt{\hat{\alpha}^\prime \over 2}\hat{p}^\mu$ represents the zero-mode part in terms of the string's center-of-mass momentum $\hat{p}^\mu$, and the constant $\hat{\alpha}^\prime$ (with dimensions of length-squared) replaces the analogous ${\alpha}^\prime$ parameter in the tensile mode expansion \eqref{MW9}. The harmonic oscillators $\lbrace \tilde{c}_n^\mu,c_n^\mu, \tilde{c}_{-n}^\mu,c_{-n}^\mu\rbrace$ correspond to the operators of right- and left-moving tensionless worldsheet modes. These operators do not interact with each other and satisfy the commutation relations,
\begin{align}\label{IL18}
	\big[\tilde{c}_n^\mu, \tilde{c}_m^\nu\big] =  \big[{c}_n^\mu, {c}_m^\nu\big] = n\eta^{\mu\nu}\delta_{n+m,0},
\end{align}
with their respective Hermitian conjugates $\tilde{c}_{-n}^\mu=(\tilde{c}_n^\mu)^\dagger$ and ${c}_{-n}^\mu=({c}_n^\mu)^\dagger$. The right- and left-moving $\lbrace\tilde{V}_n,V_n\rbrace$, along with their conjugates $\lbrace \tilde{V}_n^*, V_n^*\rbrace$ have a frequency $\omega_n = {2\pi cn \over \ell}$, are successively given by:
\begin{align}\label{IL20}
	\begin{split}
		\tilde{V}_n &= {i\left(1-i\omega_n\tau\right) e^{i\omega_n \sigma/c} \over \sqrt{4\pi}n}, \qquad \tilde{V}_n^* = -{i\left(1+i\omega_n\tau\right) e^{-i\omega_n \sigma/c} \over \sqrt{4\pi}n}, \\[10pt]
		V_n &= {i\left(1-i\omega_n\tau\right) e^{-i\omega_n \sigma/c} \over \sqrt{4\pi}n}, \qquad V_n^* = -{i\left(1+i\omega_n\tau\right) e^{i\omega_n \sigma/c} \over \sqrt{4\pi}n}.
	\end{split}
\end{align}

Next, we will explore how the $c$-oscillators intrinsically characterize the quantum structure of a tensionless worldsheet and how they map to their tensile counterparts, i.e., the $\alpha$-oscillators. To do this, we must revisit the framework developed in \cite{Bagchi:2013bga,Bagchi:2015nca}, which introduces a parametric space by limiting the tensile theory, corresponding to the string tension being exactly zero. This approach successfully recovers all the physics and results associated with intrinsic tensionless strings. The basic premise of this limiting approach lies in the physical interpretation of the tensionless limit of fundamental strings. As the tension approaches zero, fundamental strings become long and floppy, and their length tends to infinity. Technically, this situation can be realized in worldsheet coordinates by setting the limit $\sigma \to \infty$. However, directly applying $\sigma \to \infty$ would disrupt the ``closedness'' condition \eqref{MWx1} of strings, since their ends are fixed and identified in terms of a periodic $\sigma$. Instead of taking $\sigma \to \infty$, we can alternatively consider a limit where \cite{Bagchi:2015nca},
\begin{align}\label{LA1}
	\sigma \to \sigma, \quad \tau \to \epsilon \tau, \quad \alpha^\prime \equiv {\hat{\alpha}^\prime \over \epsilon}, \quad \epsilon \to 0.
\end{align}
The above approach is referred to as the \textit{worldsheet contraction}. The tensionless limit \eqref{LA1} corresponds to an infinite boost limit $\bm{\beta} = {\bm{\sigma} \over c\tau} \to \infty$ on the worldsheet. Consequently, this infinite boost tensionless limit can also be understood as the so-called \textit{Carrollian} or \textit{ultra-relativistic} limit \cite{Duval:2014uoa,deBoer:2021jej},\footnote{Since the Minkowski worldsheet is inertial, the infinite boost limit $\bm{{\beta}} = {\bm{v} \over c} \to \infty$ can be achieved only by taking the Carrollian limit $c \to 0$ for a uniform worldsheet velocity $\bm{v}$.}
\begin{align}\label{LA2}
	c \to \epsilon c, \quad \bm{{\beta}}  \equiv {\bm{\hat{\beta}} \over \epsilon}, \quad \epsilon \to 0,
\end{align}
where $\bm{\hat{\beta}}$ is the Carroll boost parameter. This algebraic realization sets the speed of light $c$ to zero on the closed string worldsheet, rendering it a null or degenerate manifold. Notably, one must adopt either limit \eqref{LA1} or \eqref{LA2} exclusively to avoid ambiguity in structuring a tensionless worldsheet. By carefully implementing the typical limit \eqref{LA1} at every stage of the tensile worldsheet setup in \cref{wmb}, one can validate the quantum mode expansion \eqref{IL19} of an intrinsic tensionless string worldsheet. This process leads to deriving the intrinsic tensionless oscillation operators in terms of the tensile $\alpha$ operators. Finally, switching to the harmonic $c$-oscillator basis precisely reproduces the mode expansion \eqref{IL19} associated with the contracted quantum modes \eqref{IL20}, resulting in the following Bogoliubov transformations between the tensionless and tensile mode creation and annihilation operators (for all $n > 0$): 
\begin{align}\label{LA5}
\begin{split}
\tensor*{\tilde{c}}{_n^\mu} &= \beta_+\tensor*{\tilde{\alpha}}{_n^\mu} + \beta_-\tensor*{{\alpha}}{_{-n}^\mu},\qquad \tensor*{{c}}{_n^\mu} = \beta_+ \tensor*{{\alpha}}{_n^\mu} + \beta_- \tensor*{\tilde{\alpha}}{_{-n}^\mu}, \\[6pt]
\tensor*{\tilde{c}}{_{-n}^\mu} &= \beta_+\tensor*{\tilde{\alpha}}{_{-n}^\mu} + \beta_-\tensor*{{\alpha}}{_{n}^\mu},\qquad \tensor*{{c}}{_{-n}^\mu} = \beta_+ \tensor*{{\alpha}}{_{-n}^\mu} + \beta_- \tensor*{\tilde{\alpha}}{_{n}^\mu},
\end{split}
\end{align}
where $\beta_{\pm} = {1 \over 2}\left(\sqrt{\epsilon} \pm {1\over \sqrt{\epsilon}}\right)$. One must keep in mind that all the above identifications remain valid as $\epsilon \to 0$. However, the current framework of the tensionless worldsheet induces a parametric evolution in $\epsilon$ that defines a natural `flow' valid throughout the parameter space $\epsilon \in [1,0]$. This flow essentially leads us to interpret the $c$-oscillators \eqref{LA5} as a set of evolving creation and annihilation operators $\lbrace\tensor*{\tilde{c}}{_n^\mu}(\epsilon),\tensor*{{c}}{_n^\mu}(\epsilon),\tensor*{\tilde{c}}{_{-n}^\mu}(\epsilon),\tensor*{{c}}{_{-n}^\mu}(\epsilon)\rbrace$, which interpolate between the original tensile pairs $\left\lbrace \tilde{\alpha}_n^\mu,{\alpha}_n^\mu,\tilde{\alpha}_{-n}^\mu,{\alpha}_{-n}^\mu\right\rbrace$ at $\epsilon = 1$ and their tensionless counterparts $\lbrace \tensor*{\tilde{c}}{_n^\mu},\tensor*{{c}}{_n^\mu},\tensor*{\tilde{c}}{_{-n}^\mu},\tensor*{{c}}{_{-n}^\mu}\rbrace$ as $\epsilon \to 0$. These operators indeed annihilate an evolving vacua $\lvert 0_{c}(\epsilon)\rangle$ , which changes with respect to the $\epsilon$ parameter: 
\begin{align}\label{LA6}
	\lvert 0_{c}(\epsilon)\rangle :~	\tensor*{\tilde{c}}{_n^\mu}(\epsilon)\lvert 0_{c}(\epsilon)\rangle = \tensor*{{c}}{_n^\mu}(\epsilon)\lvert 0_{c}(\epsilon)\rangle  = 0, \enspace \forall~ n>0.
\end{align} 
The tensionless vacuum state $\lvert 0_{c}\rangle$ is obtained as a special limiting case of the tensile evolving vacua \eqref{LA6}, i.e., $\lvert 0_{c}\rangle = \lim_{\epsilon \to 0} \lvert 0_{c}(\epsilon)\rangle$.
	
\section{Milne road to time-evolving worldsheets}\label{wkb}
This section elucidate the setup of embedding the Milne physics on string target space and worldsheets, followed by exploring their quantum mode expansion and the Bogoliubov transformations of related oscillators. 
	
\subsection{From Minkowski to Milne worldsheet: the embedding setup}\label{ETS}
Now, we aim to explore the closed string worldsheet moving in a Milne target spacetime, which resides in the analogous F and P regions as in \cref{fig:1}. In this context, we first express the target spacetime metric for the Minkowski worldsheet \eqref{MW1} as
\begin{align}\label{ES1}
		ds^2 = \left(cdX^0\right)^2 - \left(dX^1\right)^2 - \left(dX^i\right)^2, \quad \forall~ i \in (2, D-2).
\end{align}
We then consider the inertial components $\lbrace X^0, X^1 \rbrace$ subject to the standard transformations \eqref{I2} into a new accelerated set $\lbrace \bar{X}^0, \bar{X}^1 \rbrace$ while keeping the other $D-2$ spatial directions $X^i$ invariant, yielding:
\begin{align}\label{ES2}
		X^0 = \pm a^{-1}e^{a\bar{X}^0}\cosh\left({a\bar{X}^1/ c}\right), \enspace X^1 = \pm{ca^{-1}} e^{a\bar{X}^0} \sinh\left({a\bar{X}^1/ c}\right), \enspace  X^i = \bar{X}^i,
\end{align}
where $\pm$ denotes the F and P wedges, respectively. The above transformations, in turn, embed the closed string worldsheet into a time-evolving (expanding in F and contracting in P) Milne target spacetime having a proper Rindler acceleration analogous $ac$. Then the metric describing this new target spacetime is given by,
\begin{align}\label{ES3}
		ds^2 = e^{2a\bar{X}^0}\left[\left(cd\bar{X}^0\right)^2 - \left(d\bar{X}^1\right)^2\right] - \left(dX^i\right)^2,
\end{align}
where the related Milne horizons correspond to the regime $\bar{X}^0 \to -\infty$ and $\bar{X}^1 \to \pm\infty$ by setting an infinite {time-evolution} limit ${a\over c} \to \infty$. Notably, throughout this work, we always consider the closed string with a Milne worldsheet effectively propagating only in two directions, and the $i$-directions do not depend on the related worldsheet coordinates. 
	
It is also important to note the notion of two distinct observers associated with the Minkowski and Milne worldsheets: one being an inertial observer traveling through flat Minkowski spacetime, and the other being a time-evolving observer moving within Milne spacetime. The induced metric on the newly embedded worldsheet is expressed as follows
\begin{align}\label{ES4}
h_{ab} \propto e^{2a\bar{X}^0}\partial_aX^{\mu}\partial_bX^{\nu}\eta_{\mu\nu}, \quad \forall~ \mu,\nu \in (0, 1).
\end{align}
This metric remains regular everywhere except in the limit where $\bar{X}^0 \to -\infty$, which corresponds to approaching the horizons of the Milne target space. In this limit, the metric \eqref{ES4} degenerates as its conformal factor $e^{2a\bar{X}^0}$ becomes non-invertible and vanishes. In other words, when embedding a worldsheet into a Milne target space, certain `divergent points' also emerge on the worldsheet metric. These points can be interpreted as the locations where a worldsheet horizon is induced and coincides with the target space horizon. In \cref{TAW}, we will also explore how these worldsheet divergent points correspond to the limit ${a\over c} \to \infty$ where the Milne worldsheet becomes `null' or `tensionless'.
	
In this work, we examine a scenario in which the worldsheet of the time-evolving string under consideration is assumed to uphold the same physical structure as the target spacetime in which it is embedded. More precisely, we argue that when a string propagates in a Milne background, the worldsheet can inherit a corresponding Milne structure from the target spacetime.\footnote{It is important to underscore that such a choice of matching geometrical structures between the worldsheet and target spacetime is not a universal feature. There exist numerous examples in string theory—such as the Gomis–Ooguri string \cite{Gomis:2000bd}—where the worldsheet and target space exhibit different structures.} Therefore, when considering the mapping $X^\mu(\tau,\sigma) \to \bar{X}^\mu(f(\tau,\sigma))$, we must associate the Milne worldsheet with new coordinates $(\xi,\eta)$, which are reparametrizations of their Minkowski counterparts $(\tau,\sigma)$, given by
\begin{align}\label{CP1}
\begin{split}
			&{\rm (F) ~:~} 	\tau =  a^{-1}e^{a\xi}\cosh\left({a\eta/ c}\right),\quad
			\sigma = {ca^{-1}} e^{a\xi} \sinh\left({a\eta/ c}\right),\\[10pt]
			&{\rm (P) ~:~} 	\tau = -a^{-1} e^{a\xi}\cosh\left({a\eta/ c}\right),\quad
			\sigma = -{ca^{-1}} e^{a\xi} \sinh\left({a\eta/ c}\right).
\end{split}
\end{align}
These transformations induce a two-dimensional metric structure for the closed string Milne worldsheet,
\begin{align}\label{CP2}
ds^2 = e^{2a\xi}\left(c^2d\xi^2 - d\eta^2\right).
\end{align}
Here all the related trajectories describing the evolution of the Milne worldsheet are divided into the causally disconnected regions F $(c\tau > \lvert \sigma \rvert)$ and P $(c\tau < \lvert \sigma \rvert)$. These regions are bounded by two horizons, defined in terms of the worldsheet coordinates as
\begin{align}
c\tau = \pm \sigma, \enspace \eta = \pm \infty, \enspace \xi = -\infty,
\end{align}
where the worldsheet time-evolution hits the limit $a \to \infty$. At this point, we can also define the inverse mapping between the coordinates of the Minkowski and Milne worldsheets via
\begin{align}\label{CP3}
\eta = {c\over 2a}\ln \left({c\tau + \sigma \over c\tau-\sigma}\right), \quad \xi =  {1\over 2a}\ln \left({a^2 \over c^2}\left(c^2\tau^2 - \sigma^2 \right)\right).
\end{align}
Notably, this mapping diverges at the two singular points $c\tau = \pm \sigma$ where the Milne worldsheet metric \eqref{CP2} degenerates. This corresponds to the analogous notion of reaching the horizons of the associated Milne target space, which in turn induces similar horizons on its worldsheet. 

\subsection{Periodicity and its regularization}\label{CP}
However, it is important to note that the closed string worldsheet with Milne or accelerated coordinates $(\xi, \eta)$ is technically distinct from the worldsheet with Minkowski or inertial reparametrization $(\tau, \sigma)$. Arguably, the mapping from an inertial to an accelerated worldsheet actually induces ``folds'' on the associated closed strings \cite{Bagchi:2021ban}. In this context, the divergent points $c\tau = \pm \sigma$ associated with the mapping \eqref{CP3} are precisely where these folds appear. The incorporation of time-evolution on the worldsheet causes the closed string to become increasingly elongated, eventually becoming infinitely stretched at the divergent or fold points (e.g., see \cref{fig:3}). The entire scenario can be viewed as if the Milne worldsheet splits into two causally disconnected pieces in the $(\xi, \eta)$ parametrization, corresponding to $c\tau > \sigma$ (top Milne wedge F) and $c\tau < \sigma$ (down Milne wedge P). In other words, the Milne worldsheet can be interpreted as two open strings glued together at the fold points $c\tau = \pm \sigma$, with appropriate boundary conditions to preserve the closed string properties.
	
In the above discussion, we now need to explore how the periodicity of the worldsheet is affected by the Milne parametrization $(\tau, \sigma) \rightarrow \left(\xi(\tau, \sigma), \eta(\tau, \sigma) \right)$. This is crucial in defining the `closedness' condition for the string with Milne worldsheet $\bar{X}^\mu\left(\xi(\tau, \sigma), \eta(\tau, \sigma) \right)$. From the perspective of the Minkowski target space, the closed string worldsheet always satisfies the boundary condition,
\begin{align}\label{CP4}
		X^\mu (\tau, \sigma) = X^\mu (\tau + \Delta\tau, \sigma + \Delta\sigma), \enspace \Delta\tau = 0, \enspace \Delta\sigma = \ell,
\end{align}  
where $\ell$ is an arbitrary periodicity of the Minkowski worldsheet explicitly associated with the coordinate $\sigma$. Evidently, when transitioning from the Minkowski to Milne mapping $X^\mu (\tau, \sigma) \to \bar{X}^\mu \left(\xi(\tau, \sigma), \eta(\tau, \sigma) \right)$, this condition \eqref{CP4} must be modified. For the closed string to remain closed in this new embedding, we must impose the following condition for $\mu = (0,1)$:
\begin{eqnarray}\label{CP5}
&X^\mu (\tau, \sigma) =\left\{
\begin{array}{lc}\nonumber
\pm a^{-1}e^{a\bar{X}^0\left(\xi(\tau,\sigma), \eta(\tau,\sigma)\right)}\cosh\left({a\bar{X}^1\left(\xi(\tau,\sigma), \eta(\tau,\sigma)\right)/ c}\right) \\~\\
\pm{ca^{-1}} e^{a\bar{X}^0\left(\xi(\tau,\sigma), \eta(\tau,\sigma)\right)} \sinh\left({a\bar{X}^1\left(\xi(\tau,\sigma), \eta(\tau,\sigma)\right)/ c}\right)
\end{array} \right., \\~\\[6pt]
&= X^\mu (\tau, \sigma + \ell) =\left\{
\begin{array}{lc}\nonumber
\pm a^{-1}e^{a\bar{X}^0\left(\xi(\tau,\sigma+\ell), \eta(\tau,\sigma+\ell)\right)}\cosh\left({a\bar{X}^1\left(\xi(\tau,\sigma+\ell), \eta(\tau,\sigma+\ell)\right)/ c}\right)  \\~\\
\pm{ca^{-1}} e^{a\bar{X}^0\left(\xi(\tau,\sigma+\ell), \eta(\tau,\sigma+\ell)\right)} \sinh\left({a\bar{X}^1\left(\xi(\tau,\sigma+\ell), \eta(\tau,\sigma+\ell)\right)/ c}\right)
\end{array} \right..
\end{eqnarray}
With this periodic condition in the embedding of the non-inertial target space, the related Milne worldsheet with $\left(\xi, \eta \right)$ parametrization naively appears to inherit effective periodicity components along both the spatial $\eta$ and temporal $\xi$ directions, such that
\begin{subequations}\label{CP6}
\begin{align}\label{CPYY}
		\bar{X}^\mu (\xi, \eta) = \bar{X}^\mu (\xi + \Delta\xi, \eta + \Delta\eta).
\end{align} 
Technically, the periodicity shifts $\Delta\xi$ and $\Delta\eta$ on the Milne worldsheet $\bar{X}^\mu (\xi, \eta)$ originate from the Minkowski worldsheet periodicity $\Delta\sigma = \ell$ in \eqref{CP4} through the parametrization $(\tau, \sigma) \rightarrow \left(\xi(\tau, \sigma), \eta(\tau, \sigma) \right)$. However, it remains to be verified whether this original Minkowski periodicity genuinely induces a two-component periodic structure through independent shifts along both $\eta$ and $\xi$, or whether the Milne parametrization ultimately selects only one of these components. For that, we must utilize the inverse Milne coordinate mappings \eqref{CP3} and express $\Delta\xi$ and $\Delta\eta$ in terms of the Minkowski worldsheet coordinates $(\tau, \sigma)$. It is also crucial to note that the equivalence between the `closedness' conditions in \cref{CP4,CPYY} is well-defined and comparable only at $\tau = \eta = 0$.\footnote{This can be understood by considering the Milne worldsheet as a hyperboloid with constant $\eta$ slices forming elliptical cross-sections. In the inertial case, at constant $\tau$, smoothly connected circular slices appear parallel to the $x$-axis. The origins of the inertial $\tau$ and non-inertial $\eta$ slices coincide only at $\tau = \eta = 0$.} Consequently, to properly isolate the effect of the original periodicity $\Delta\sigma = \ell$ in a manner consistent with both coordinate descriptions, it is convenient to evaluate these quantities at the common reference point $\tau = \eta = 0$. This procedure defines $\Delta\xi = \xi(0,\ell) - \xi(0,0)$ and $\Delta\eta = \eta(0,\ell) - \eta(0,0)$, ultimately leading to
\begin{align}\label{CPXX}
\begin{gathered}
\Delta\xi = {1\over 2a}\left[\ln \left(-{a^2\ell^2 \over c^2}\right)-\ln \left(0\right)\right] \equiv  \mathcal{L}_{\rm eff},\quad \Delta\eta = 0.
\end{gathered}
\end{align}	
\end{subequations}     
Interestingly, the above analysis shows that only the timelike shift $\Delta\xi$ survives, causing the original Minkowski closed-string condition $\sigma \sim \sigma+\ell$ to re-emerge as the effective periodicity $\mathcal{L}_{\rm eff}$ along $\xi \sim \xi + \mathcal{L}_{\rm eff}$ on the Milne worldsheet. To properly interpret this, let us revisit the transformation \eqref{CP5}, which shifts the worldsheet target space from Minkowski to Milne. This transformation introduces a non-inertial frame evolving in time by setting the Milne parameter $a$ along the timelike Minkowski target-space direction $X^0$. Consequently, this non-inertial time evolution is inherited by the Milne worldsheet itself, causing its geometry to deform through an effective stretching along the timelike coordinate $\xi$. To counterbalance this deformation, the closed Milne worldsheet must incorporate the periodicity condition \eqref{CP6}. Notably, this mechanism is qualitatively distinct from the corresponding construction of accelerated Rindler worldsheets developed in \cite{Bagchi:2021ban} (see, for example, Section 5.2 therein). In the Rindler case, the non-inertial frame is generated by setting the acceleration parameter ($a^\prime$) along the spacelike Minkowski target-space direction $X^1$. Consequently, the resulting deformation is inherited by the Rindler worldsheet through stretching along its spacelike coordinate, thereby generating an effective spacelike periodicity. In this sense, the Rindler construction preserves the spacelike character of the original Minkowski closed-string periodicity. By contrast, periodicity on the Milne worldsheet acquires a fundamentally different character through its realization along the timelike coordinate $\xi$.

However, the above setting also induces a singularity in the computation of the effective periodicity $\mathcal{L}_{\rm eff}$. In other words, $\mathcal{L}_{\rm eff}\rightarrow\infty$ in \eqref{CPXX}, which is an artifact of the divergence associated with the worldsheet horizons situated at $c\tau=\pm\sigma$. To regulate this, we require that the closedness condition \eqref{CP6} be accompanied by an effective periodicity of the following form:
\begin{align}\label{CP7}
		\mathcal{L}_{\rm eff}  = \lim_{\varphi \to 0} {1\over 2a}\left[\ln \left({a^2 \over c^2}\left(0 - (\varphi + \ell)^2 \right)\right)-\ln \left({a^2 \over c^2}\left(0 - \varphi^2 \right)\right)\right].
\end{align} 
In this simplistic setup, we set a constant parameter $\varphi$ as the initial value of the inertial $\sigma$ coordinate and define a regularized effective periodicity $\mathcal{L}_{\varphi}$ for the Milne worldsheet of closed strings, given as
\begin{align}\label{CP8}
		\mathcal{L}_{\varphi} = {1\over a}\ln \left(1 + {\ell\over \varphi}\right) =  \xi(0,\varphi + \ell) - \xi(0,\varphi).
\end{align} 
Further, the above regularization setup needs to be appropriately accounted for in the original mapping \eqref{CP1} between Minkowski and Milne worldsheets. The regularized transformations of their coordinates are given by
\begin{align}\label{CP9}
\begin{split}
&{\rm (F) ~:~} 	\tau =  a^{-1}e^{a(\xi - \psi)}\cosh\left({a\eta/ c}\right),\quad
\sigma + \varphi = {ca^{-1}} e^{a(\xi - \psi)} \sinh\left({a\eta/ c}\right),\\[6pt]
&{\rm (P) ~:~} 	\tau = -a^{-1} e^{a(\xi - \psi)}\cosh\left({a\eta/ c}\right),\quad
\sigma + \varphi = -{ca^{-1}} e^{a(\xi - \psi)} \sinh\left({a\eta/ c}\right),
\end{split}
\end{align}
where $\psi$ is introduced as a regularization parameter in the Milne worldsheet coordinates to capture the effect of considering the initial value $\xi(0,\varphi)$ in step \eqref{CP8}. Now, if we look at the two coordinate patches $(\tau,\sigma)$ and $(\xi,\eta)$ in the regularized mapping \eqref{CP9}, there exists an offset between their origins. However, this mismatch can be adjusted by tuning the non-inertial parameter $\psi$ in terms of the inertial parameter $\varphi$ via,
\begin{align}\label{CP10}
		\varphi =\pm{c\over a}e^{-a\left(\psi - {i\pi \over 2a}\right)}. 
\end{align}
It is evident that as we approach the horizon $c\tau = \pm \sigma$ of the non-inertial worldsheet, its time-evolution increases (i.e., ${a\over c} \to \infty$), causing the regularization parameter $\varphi \to 0$ while assuming $\psi$ is positive.\footnote{To ensure that $\psi$ remains positive in the definition \eqref{CP10}, it is necessary to satisfy the condition $0 < \lvert {a\varphi \over c} \rvert < 1$. In the present analysis, we assume this inequality holds consistently, even as we approach the Milne worldsheet horizon limit ${a \over c} \to \infty$, by carefully fixing the regularization parameter $\varphi \to 0$.} In this extremal limit, the effective periodicity approaches:
\begin{align}\label{X1}
		\mathcal{L}_{\rm eff} \to \mathcal{L}_{\varphi} \approx {1\over a}\ln \left({a\ell\over c} \right) + \psi.
\end{align}
In particular, within the present regularization setup, the horizon-limit $a/c \to \infty$ responsible for the divergence of the Milne worldsheet periodicity simultaneously drives the regulating parameter to vanish in such a way that a finite periodicity survives. The same limit also degenerates the induced Milne worldsheet metric, as discussed in \cref{ETS}. We would like to emphasize that such degeneration of the worldsheet geometry into a Carrollian structure is an intrinsic feature of the tensionless string phase. A strong consistency check appears from the fact that the regularized periodicity \eqref{X1} is further shown in \cref{TAW} to generate a tensionless sector for the Milne worldsheet. In this sense, the regularization procedure employed here is geometrically natural and physically unambiguous: it consistently extracts the finite contribution from the divergent worldsheet periodicity structure while preserving the underlying null horizon structure and the associated emergent tensionless sector of the worldsheet. Remarkably, this regularization prescription also provides a convenient framework for handling similar divergences arising in such non-inertial worldsheet periodicities, and further discussions may be found in \cite{Bagchi:2021ban}.

At this point, the inverse transformations between the regularized worldsheet coordinates are expressed as
\begin{align}\label{CP11}
\eta = {c\over 2a}\ln \left({c\tau + \sigma + \varphi\over c\tau-\sigma - \varphi}\right), \quad \xi - \psi =  {1\over 2a}\ln \left({a^2 \over c^2}\left(c^2\tau^2 - (\sigma^2 + \varphi)^2 \right)\right).
\end{align}
The above-developed setup of regularized coordinates and periodicity of the Milne worldsheet will play a crucial role in the analysis of their quantum mode expansion in the next section. Notably, readers familiar with the literature may find these relations closely analogous to those of the Rindler worldsheet \cite{Bagchi:2020ats,Bagchi:2021ban}, although the interpretation of the parameters involved is entirely different.
	
	
\subsection{Quantum mode expansions}\label{KME}
We have already established that the two-dimensional metric structures of the Minkowski and Milne worldsheets are conformally connected. Therefore, the Milne worldsheet $\bar{X}^\mu(\xi, \eta)$ must satisfy the same two-dimensional massless Klein-Gordon equations of motion (EOM) as the conformal gauge-fixed Minkowski worldsheet $X^\mu (\tau, \sigma)$:
	\begin{align}\label{KME1}
	\ \Box_{\xi,\eta} \bar{X}^{\mu} = 0 =\Box_{\tau,\sigma} X^{\mu},
	\end{align}
along with the Virasoro constraints:
\begin{align}
	\partial_\xi \bar{X}^\mu\partial_\eta \bar{X}_\mu = 0, \quad (\partial_\xi \bar{X}^\mu)^2 + (\partial_\eta \bar{X}^\mu)^2 = 0,
\end{align} 
where $\Box_{\xi,\eta} = (c^{-2} \partial_\xi^2 - \partial_\eta^2)$ is the D'Alembertian on the worldsheet with $(\xi, \eta)$ reparametrization. Consequently, we can now express the corresponding mode expansion solution, following the schematic form of \eqref{MW9} for the inertial worldsheet. Here, each mode-expansion component must be expanded with the associated mode functions and oscillation operators in both the F and P wedges for the Milne worldsheet. In this context, the most general solution to the EOM \eqref{KME1} is given by:     
\begin{align}\label{KME2}
	\bar{X}^\mu (\xi, \eta) &= \bar{x}^\mu + \alpha^\prime \bar{p}^\mu\xi + \sqrt{2\pi\alpha^\prime}\sum_{\substack{n > 0; \\ \Lambda = \rm F,P}} \left[ \tensor*[^{\rm \Lambda}]{\tilde{\beta}}{_n^\mu} \tensor*[^{\rm \Lambda}]{\tilde{\mathcal{U}}}{_n} +  \tensor*[^{\rm \Lambda}]{{\beta}}{_n^\mu}\tensor*[^{\rm \Lambda}]{\mathcal{U}}{_n} + \tensor*[^{\rm \Lambda}]{\tilde{\beta}}{_{-n}^\mu} \tensor*[^{\rm \Lambda}]{\tilde{\mathcal{U}}}{_n^*} + \tensor*[^{\rm \Lambda}]{{\beta}}{_{-n}^\mu} \tensor*[^{\rm \Lambda}]{\mathcal{U}}{_n^*}\right]. \nonumber\\[6pt]
	&= \bar{x}^\mu + \alpha^\prime \bar{p}^\mu\xi + \sqrt{2\pi\alpha^\prime}\sum_{n > 0} \left[ \tensor*[^{\rm F}]{\tilde{\beta}}{_n^\mu} \tensor*[^{\rm F}]{\tilde{\mathcal{U}}}{_n}+ \tensor*[^{\rm P}]{\tilde{\beta}}{_n^\mu} \tensor*[^{\rm P}]{\tilde{\mathcal{U}}}{_n} +  \tensor*[^{\rm F}]{{\beta}}{_n^\mu}\tensor*[^{\rm F}]{\mathcal{U}}{_n} +  \tensor*[^{\rm P}]{{\beta}}{_n^\mu}\tensor*[^{\rm P}]{\mathcal{U}}{_n} + \rm h.c.\right],
\end{align}
where the center of mass parameters $\bar{x}^\mu$ and $\bar{p}^\mu$ correspond to the zero modes of the worldsheet. The coefficients $\big\lbrace \tensor*[^{\rm F}]{\tilde{\beta}}{_n^\mu}, \tensor*[^{\rm P}]{\tilde{\beta}}{_n^\mu}\big\rbrace$ and $\big\lbrace \tensor*[^{\rm F}]{{\beta}}{_n^\mu}, \tensor*[^{\rm P}]{{\beta}}{_n^\mu}\big\rbrace$ are creation operators (for $n < 0$) and annihilation operators (for $n > 0$) associated with the right- and left-moving Milne mode functions $\big\lbrace \tensor*[^{\rm F}]{\tilde{\mathcal{U}}}{_n}, \tensor*[^{\rm P}]{\tilde{\mathcal{U}}}{_n}\big\rbrace$ and $\big\lbrace \tensor*[^{\rm F}]{{\mathcal{U}}}{_n}, \tensor*[^{\rm P}]{{\mathcal{U}}}{_n}\big\rbrace$. These oscillators satisfy the commutation relations,
\begin{align}\label{KME3}
\big[\tensor*[^{\rm \Lambda}]{\tilde{\beta}}{_n^\mu},\tensor*[^{\rm {\Lambda^\prime}}]{\tilde{\beta}}{_m^\nu}\big] = \big[\tensor*[^{\rm \Lambda}]{\beta}{_n^\mu},\tensor*[^{\rm {\Lambda^\prime}}]{\beta}{_m^\nu}\big] = n\eta^{\mu\nu}\delta_{\Lambda,\Lambda^\prime}\delta_{n+m,0}, \enspace \forall~ \Lambda,\Lambda^\prime = \rm (F,P)
\end{align} 
and are related via their respective Hermitian conjugates $\tensor*[^{\rm \Lambda}]{\tilde{\beta}}{_{-n}^\mu}=\big(\tensor*[^{\rm \Lambda}]{\tilde{\beta}}{_n^\mu}\big)^\dagger$ and $\tensor*[^{\rm \Lambda}]{{\beta}}{_{-n}^\mu}=\big(\tensor*[^{\rm \Lambda}]{{\beta}}{_n^\mu}\big)^\dagger$. Furthermore, they annihilate the Milne worldsheet vacuum state $\lvert0_{\rm K}\rangle$, defined to obey:
\begin{align}\label{KME4}
\tensor*[^{\rm F}]{\tilde{\beta}}{_n^\mu}\lvert0_{\rm K}\rangle = \tensor*[^{\rm P}]{\tilde{\beta}}{_n^\mu}\lvert0_{\rm K}\rangle = \tensor*[^{\rm F}]{{\beta}}{_n^\mu}\lvert0_{\rm K}\rangle = \tensor*[^{\rm P}]{{\beta}}{_n^\mu}\lvert0_{\rm K}\rangle = 0, \enspace \forall~ n>0.
\end{align}
We now turn to a critical aspect of this analysis: the form of the mode expansion functions that comprise the quantized Milne worldsheet solution \eqref{KME2}. These functions are assumed to have a structure similar to their Minkowski counterparts in \eqref{MW10}. Within the context of the regularized $(\xi, \eta)$ mapping depicted in \cref{CP9,CP11}, Milne mode functions must be associated with a redefined and regulated frequency,
\begin{align}\label{KME5}
		\Omega_{n} = {2\pi cn \over \mathcal{L}_{\varphi}},
\end{align}
where $\mathcal{L}_{\varphi}$ is the regularized periodicity \eqref{CP8} along the $\xi$-coordinate of the Milne worldsheet. We now introduce the light-cone coordinates valid throughout the Minkowski worldsheet:
\begin{align}\label{KME6}
\sigma^+ = c\tau + \sigma, \quad \sigma^- = c\tau - \sigma,
\end{align}
and their analogs in the Milne worldsheet F and P wedges are:
\begin{align}\label{KME7}
\begin{split}
&{\rm (F) ~:~} 	\eta^+_{\rm F} = c\xi + \eta,\quad \eta^-_{\rm F} = c\xi - \eta, \\[6pt]
&{\rm (P) ~:~} 	\eta^+_{\rm P} = -c\xi + \eta,\quad \eta^-_{\rm P} = -c\xi - \eta.
\end{split}
\end{align}
Within our regularization scheme, these inertial and non-inertial sets are related as follows
\begin{eqnarray}\label{KME8}
\sigma^+ + \varphi =\left\{
\begin{array}{lc}
+{c\over a}e^{{a\over c}\left(\eta^+_{\rm F} - c\psi\right)} & : \ \rm F \\~\\
-{c\over a}e^{-{a\over c}\left(\eta^-_{\rm P} + c\psi\right)} & : \ \rm P
\end{array} \right.,
~~~~~~
\sigma^- - \varphi =\left\{
\begin{array}{lc}
+{c\over a}e^{{a\over c}\left(\eta^-_{\rm F} - c\psi\right)} & : \ \rm F \\~\\
-{c\over a}e^{-{a\over c}\left(\eta^+_{\rm P} + c\psi\right)} & : \ \rm P
\end{array} \right..
\end{eqnarray}
In terms of the above light-cone coordinates, the right-moving (along $\eta^-_{\rm F}$  and $\eta^-_{\rm P}$) and left-moving (along $\eta^+_{\rm F}$  and $\eta^+_{\rm P}$) mode functions of frequency $\Omega_{n}$, as well as their Hermitian conjugates in the F and P wedges associated with the quantized Milne worldsheet, can be reexpressed as follows:
\begin{subequations}\label{KME9}
\allowdisplaybreaks{\begin{align}
\tensor*[^{\rm F}]{\tilde{\mathcal{U}}}{_n} &= {i e^{-i\Omega_{n} \eta^-_{\rm F}/c} \over \sqrt{4\pi}n}, \qquad \tensor*[^{\rm F}]{\tilde{\mathcal{U}}}{_n^*} = -{i e^{i\Omega_{n} \eta^-_{\rm F}/c} \over \sqrt{4\pi}n},\label{KME9a} \\[6pt]
\tensor*[^{\rm P}]{\tilde{\mathcal{U}}}{_n} &= {i e^{-i\Omega_{n} \eta^-_{\rm P}/c} \over \sqrt{4\pi}n}, \qquad \tensor*[^{\rm P}]{\tilde{\mathcal{U}}}{_n^*} = -{i e^{i\Omega_{n} \eta^-_{\rm P}/c} \over \sqrt{4\pi}n},\label{KME9b} \\[6pt]
\tensor*[^{\rm F}]{{\mathcal{U}}}{_n} &= {i e^{-i\Omega_{n} \eta^+_{\rm F}/c} \over \sqrt{4\pi}n}, \qquad \tensor*[^{\rm F}]{{\mathcal{U}}}{_n^*} = -{i e^{i\Omega_{n} \eta^+_{\rm F}/c} \over \sqrt{4\pi}n}, \label{KME9c}\\[6pt]
\tensor*[^{\rm P}]{{\mathcal{U}}}{_n} &= {i e^{-i\Omega_{n} \eta^+_{\rm P}/c} \over \sqrt{4\pi}n}, \qquad \tensor*[^{\rm P}]{{\mathcal{U}}}{_n^*} = -{i e^{i\Omega_{n} \eta^+_{\rm P}/c} \over \sqrt{4\pi}n}.\label{KME9d}
\end{align}}
\end{subequations}
	
\subsection{Bogoliubov transformations}\label{KBT}
To better explore the Milne worldsheet from the perspective of the Minkowski worldsheet, we aim to derive the relevant Bogoliubov transformations that connect their quantum modes and oscillation operators. For this purpose, we follow the standard and elegant method developed by Unruh \cite{Unruh:1976db}. The process begins by considering the modes of the Milne worldsheet that are analytically valid throughout the entire spacetime, including both the F and P wedges. However, it is important to note that the original or local Milne modes \eqref{KME9} are non-analytic, as $\tensor*[^{\rm F}]{\tilde{\mathcal{U}}}{_n} = \tensor*[^{\rm F}]{{\mathcal{U}}}{_n} = 0$ in P and $\tensor*[^{\rm P}]{\tilde{\mathcal{U}}}{_n} = \tensor*[^{\rm P}]{{\mathcal{U}}}{_n} = 0$ in F. To resolve this, we need to restructure these local mode functions into appropriate linear combinations that allow for analytic continuation between the F and P wedges. In this analysis, we introduce a novel basis in which these new combinations of modes as the ``global modes'' for the Milne worldsheet at the Minkowski level. These are the so-called Unruh modes \cite{Unruh:1976db}, which analytically match the natural positive-frequency Minkowski modes in the process of transitioning from Minkowski to the Milne (or more generally, any non-inertial) frame.

The detailed derivation of the global modes for the Milne worldsheet is not our primary focus here. Nevertheless, we have outlined the essential steps in \cref{FGM} for readers interested in the intricacies of the process. The typical expressions presented in \cref{KBT2,KBT3,KBT5,KBT6} are unnormalized. To properly normalize these right- and left-moving global Milne modes in the F and P wedges, we determine the appropriate normalization constants to make them orthonormal. Without delving into intermediate technicalities, we present the final normalized forms of the global Milne modes as follows
\begin{subequations}\label{KBT16}
\allowdisplaybreaks{\begin{align}
\tensor*[^{\rm (1)}]{\tilde{\mathcal{W}}}{_n} &=	{1 \over \sqrt{2\sinh\left(\pi \Omega_{n}\over a\right)}}\left(e^{\pi \Omega_{n} \over 2a}	\tensor*[^{\rm F}]{\tilde{\mathcal{U}}}{_n} - e^{-{\pi \Omega_{n} \over 2a}}\,\tensor*[^{\rm P}]{{\mathcal{U}}}{_n^*}\right), \label{KBT16a}\\[6pt]
\tensor*[^{\rm (2)}]{\tilde{\mathcal{W}}}{_n} &=	{1 \over \sqrt{2\sinh\left(\pi \Omega_{n}\over a\right)}}\left(e^{\pi \Omega_{n} \over 2a} \tensor*[^{\rm P}]{\tilde{\mathcal{U}}}{_n} - e^{-{\pi \Omega_{n} \over 2a}}\,\tensor*[^{\rm F}]{{\mathcal{U}}}{_n^*}\right),\label{KBT16b} \\[6pt]
\tensor*[^{\rm (1)}]{{\mathcal{W}}}{_n} &= {1 \over \sqrt{2\sinh\left(\pi \Omega_{n}\over a\right)}}\left(e^{\pi \Omega_{n} \over 2a} \tensor*[^{\rm F}]{{\mathcal{U}}}{_n} - e^{-{\pi \Omega_{n} \over 2a}}\,\tensor*[^{\rm P}]{\tilde{\mathcal{U}}}{_n^*}\right),\label{KBT16c}\\[6pt]
\tensor*[^{\rm (2)}]{{\mathcal{W}}}{_n} &= {1 \over \sqrt{2\sinh\left(\pi \Omega_{n}\over a\right)}}\left(e^{\pi \Omega_{n} \over 2a}\tensor*[^{\rm P}]{{\mathcal{U}}}{_n} - e^{-{\pi \Omega_{n} \over 2a}}\,\tensor*[^{\rm F}]{\tilde{\mathcal{U}}}{_n^*}\right). \label{KBT16d}
\end{align}}
\end{subequations} 
To derive the normalized expressions above, we establish the following orthonormal relations for the local Milne worldsheet modes that form their global counterparts:   
\begin{align}\label{KBT14}
\begin{split}
\big(\tensor*[^{\rm \Lambda}]{\tilde{\mathcal{U}}}{_m},\tensor*[^{\rm \Lambda}]{\tilde{\mathcal{U}}}{_n}\big) &= \left(\tensor*[^{\rm \Lambda}]{{\mathcal{U}}}{_m},\tensor*[^{\rm \Lambda}]{{\mathcal{U}}}{_n}\right) = \delta_{m,n}, \\[6pt]
\big(\tensor*[^{\rm \Lambda}]{\tilde{\mathcal{U}}}{_m^*},\tensor*[^{\rm \Lambda}]{\tilde{\mathcal{U}}}{_n^*}\big) &= \left(\tensor*[^{\rm \Lambda}]{{\mathcal{U}}}{_m^*},\tensor*[^{\rm \Lambda}]{{\mathcal{U}}}{_n^*}\right) = - \delta_{m,n}, \qquad \forall~ \Lambda = \rm F,P \\[6pt]
\big(\tensor*[^{\rm \Lambda}]{\tilde{\mathcal{U}}}{_m},\tensor*[^{\rm \Lambda}]{\tilde{\mathcal{U}}}{_n^*}\big) &= \left(\tensor*[^{\rm \Lambda}]{{\mathcal{U}}}{_m},\tensor*[^{\rm \Lambda}]{{\mathcal{U}}}{_n^*}\right) = 0.
\end{split}
\end{align}
The related inner products in the F and P regions, which we have defined in the following manner, are expressed as
\begin{align}
	\left(\tensor*[^{\rm F}]{{\Phi}}{_1},\tensor*[^{\rm F}]{{\Phi}}{_2}\right) = -i\int d\eta \Big[\tensor*[^{\rm F}]{{\Phi}}{_1}c^{-1}\partial_\xi\tensor*[^{\rm F}]{{\Phi}}{_2^*} - \tensor*[^{\rm F}]{{\Phi}}{_2^*}c^{-1}\partial_\xi\tensor*[^{\rm F}]{{\Phi}}{_1}\Big],\label{KBT12} \\[6pt]
	\left(\tensor*[^{\rm P}]{{\Phi}}{_1},\tensor*[^{\rm P}]{{\Phi}}{_2}\right) = i\int d\eta \Big[\tensor*[^{\rm P}]{{\Phi}}{_1}c^{-1}\partial_\xi\tensor*[^{\rm P}]{{\Phi}}{_2^*} - \tensor*[^{\rm P}]{{\Phi}}{_2^*}c^{-1}\partial_\xi\tensor*[^{\rm P}]{{\Phi}}{_1}\Big],\label{KBT13}
\end{align}
where $\tensor*[^{\rm \Lambda}]{{\Phi}}{_1}$ and $\tensor*[^{\rm \Lambda}]{{\Phi}}{_2}$ are arbitrary modes in the F and P regions of the Milne worldsheet. Similarly, we can express the normalized conjugate counterparts of the global Milne worldsheet modes as
\begin{subequations}\label{KBT15}
\allowdisplaybreaks{\begin{align}
\tensor*[^{\rm (1)}]{\tilde{\mathcal{W}}}{_n^*} &= {1 \over \sqrt{2\sinh\left(\pi \Omega_{n}\over a\right)}}\left(e^{\pi \Omega_{n} \over 2a}\tensor*[^{\rm F}]{\tilde{\mathcal{U}}}{_n^*} - e^{-{\pi \Omega_{n} \over 2a}}\,\tensor*[^{\rm P}]{{\mathcal{U}}}{_n}\right), \label{KBT16e}\\[6pt]
\tensor*[^{\rm (2)}]{\tilde{\mathcal{W}}}{_n^*} &= {1 \over \sqrt{2\sinh\left(\pi \Omega_{n}\over a\right)}}\left(e^{\pi \Omega_{n} \over 2a}\tensor*[^{\rm P}]{\tilde{\mathcal{U}}}{_n^*} - e^{-{\pi \Omega_{n} \over 2a}}\,\tensor*[^{\rm F}]{{\mathcal{U}}}{_n}\right),\label{KBT16f} \\[6pt]
\tensor*[^{\rm (1)}]{{\mathcal{W}}}{_n^*} &= {1 \over \sqrt{2\sinh\left(\pi \Omega_{n}\over a\right)}}\left(e^{\pi \Omega_{n} \over 2a}\tensor*[^{\rm F}]{{\mathcal{U}}}{_n^*} - e^{-{\pi \Omega_{n} \over 2a}}\,\tensor*[^{\rm P}]{\tilde{\mathcal{U}}}{_n}\right),\label{KBT16g}\\[6pt]
\tensor*[^{\rm (2)}]{{\mathcal{W}}}{_n^*} &= {1 \over \sqrt{2\sinh\left(\pi \Omega_{n}\over a\right)}}\left(e^{\pi \Omega_{n} \over 2a}\tensor*[^{\rm P}]{{\mathcal{U}}}{_n^*} - e^{-{\pi \Omega_{n} \over 2a}}\,\tensor*[^{\rm F}]{\tilde{\mathcal{U}}}{_n}\right). \label{KBT16h}
\end{align}}  
\end{subequations}
Given that the global mode functions \eqref{KBT16} and their conjugates \eqref{KBT15} are well-defined across the entire closed string worldsheet, they can be considered as an independent basis for expanding the Milne worldsheet $\bar{X}^\mu(\xi,\eta)$. This leads to the reexpression of \eqref{KME2} in the following form:
\allowdisplaybreaks{\begin{align}\label{KBT17}
&\bar{X}^\mu (\xi, \eta) = \bar{x}^\mu + \alpha^\prime \bar{p}^\mu\xi + \sqrt{2\pi\alpha^\prime}\sum_{\substack{n > 0; \\ i = 1,2}} \left[ \tensor*[^{ (i)}]{\tilde{\gamma}}{_n^\mu} \tensor*[^{ (i)}]{\tilde{\mathcal{W}}}{_n} +  \tensor*[^{(i)}]{{\gamma}}{_n^\mu}\tensor*[^{(i)}]{\mathcal{W}}{_n} + \tensor*[^{(i)}]{\tilde{\gamma}}{_{-n}^\mu} \tensor*[^{(i)}]{\tilde{\mathcal{W}}}{_n^*} + \tensor*[^{(i)}]{{\gamma}}{_{-n}^\mu} \tensor*[^{(i)}]{\mathcal{W}}{_n^*}\right], \nonumber\\[6pt]
&= \bar{x}^\mu + \alpha^\prime \bar{p}^\mu\xi + \sqrt{2\pi\alpha^\prime}\sum_{n > 0} \left[ \tensor*[^{(1)}]{\tilde{\gamma}}{_n^\mu} \tensor*[^{(1)}]{\tilde{\mathcal{W}}}{_n}+ \tensor*[^{(2)}]{\tilde{\gamma}}{_n^\mu} \tensor*[^{(2)}]{\tilde{\mathcal{W}}}{_n} +  \tensor*[^{(1)}]{{\gamma}}{_n^\mu}\tensor*[^{(1)}]{\mathcal{W}}{_n} +  \tensor*[^{(2)}]{{\gamma}}{_n^\mu}\tensor*[^{(2)}]{\mathcal{W}}{_n} + \rm h.c.\right].
\end{align}}
Here, the global mode or Unruh oscillators $\big\lbrace \tensor*[^{(1)}]{\tilde{\gamma}}{_n^\mu}, \tensor*[^{(2)}]{\tilde{\gamma}}{_n^\mu}\big\rbrace$ and $\big\lbrace \tensor*[^{(1)}]{{\gamma}}{_n^\mu}, \tensor*[^{(2)}]{{\gamma}}{_n^\mu}\big\rbrace$ are similar to the creation operators (for $n < 0$) and annihilation operators (for $n > 0$) associated with the right- and left-moving global Milne mode functions $\big\lbrace \tensor*[^{(1)}]{\tilde{\mathcal{W}}}{_n}, \tensor*[^{(2)}]{\tilde{\mathcal{W}}}{_n}\big\rbrace$ and $\big\lbrace \tensor*[^{(1)}]{{\mathcal{W}}}{_n}, \tensor*[^{(2)}]{{\mathcal{W}}}{_n}\big\rbrace$. Since the global Milne worldsheet modes satisfy the same analyticity property as the original Minkowski modes \eqref{MW10}, they must share a common worldsheet vacuum state $\lvert0_{\rm M}\rangle$, defined to obey:
\begin{align}\label{KBT18}
\tensor*[^{(1)}]{\tilde{\gamma}}{_n^\mu}\lvert0_{\rm M}\rangle = \tensor*[^{(2)}]{\tilde{\gamma}}{_n^\mu}\lvert0_{\rm M}\rangle = \tensor*[^{(1)}]{{\gamma}}{_n^\mu}\lvert0_{\rm M}\rangle = \tensor*[^{(2)}]{{\gamma}}{_n^\mu}\lvert0_{\rm M}\rangle = 0, \enspace \forall~ n>0.
\end{align}
As a result, the global mode Milne operators $\gamma$ are generally expressed as some specific linear combinations of the Minkowski worldsheet operators $\alpha$ (and vice versa), defining the same vacuum state as described in \eqref{MW8}. Technically speaking, each right- or left-moving component of the $\alpha_n$ operator corresponds to an inter-mode mixing between $\gamma_n$ and $\gamma_{-n}$ operators associated with the same direction of propagation. Notably, the global mode operators are related to their local counterparts $\beta$ through the following inner products in the expansion \eqref{KME2}:
\begin{align}\label{KBT19}
\tensor*[^{\rm \Lambda}]{\tilde{\beta}}{_n^\mu} = \big(\bar{X}^\mu,\tensor*[^{\rm \Lambda}]{\tilde{\mathcal{U}}}{_n}\big), \quad \tensor*[^{\rm \Lambda}]{{\beta}}{_n^\mu} = \big(\bar{X}^\mu,\tensor*[^{\rm \Lambda}]{{\mathcal{U}}}{_n}\big), \qquad \forall~ n> 0,\enspace \Lambda = \rm F,P
\end{align}
To evaluate the above inner products, we first expand the Milne worldsheet $\bar{X}^\mu(\xi,\eta)$ using equation \eqref{KBT17}. Subsequently, the associated global modes are decomposed into local modes via the definitions \eqref{KBT16}, with the orthonormality identities \eqref{KBT14} and \eqref{KBT15} applied. This procedure leads to the following relationships between the non-inertial and inertial annihilation operators:
\begin{subequations}\label{KBT20}
\allowdisplaybreaks{\begin{align}
\tensor*[^{\rm F}]{\tilde{\beta}}{_n^\mu} &=  \big(\tensor*[^{(1)}]{\tilde{\mathcal{W}}}{_n},\tensor*[^{\rm F}]{\tilde{\mathcal{U}}}{_n}\big)\tensor*[^{\rm (1)}]{\tilde{\gamma}}{_n^\mu} + \big(\tensor*[^{(2)}]{{\mathcal{W}}}{_n^*},\tensor*[^{\rm F}]{\tilde{\mathcal{U}}}{_n}\big)\tensor*[^{\rm (2)}]{{\gamma}}{_{-n}^\mu} \nonumber \\[4pt]
&=	{1 \over \sqrt{2\sinh\left(\pi \Omega_{n}\over a\right)}}\left[e^{\pi \Omega_{n} \over 2a}\,	\tensor*[^{(1)}]{\tilde{\gamma}}{_n^\mu} - e^{-{\pi \Omega_{n} \over 2a}}\,\tensor*[^{(2)}]{{\gamma}}{_{-n}^\mu}\right],\\[8pt]
\tensor*[^{\rm P}]{\tilde{\beta}}{_n^\mu} &=  \big(\tensor*[^{\rm (2)}]{\tilde{\mathcal{W}}}{_n},\tensor*[^{\rm P}]{\tilde{\mathcal{U}}}{_n}\big)\tensor*[^{\rm (2)}]{\tilde{\gamma}}{_n^\mu} + \big(\tensor*[^{ (1)}]{{\mathcal{W}}}{_n^*},\tensor*[^{\rm P}]{\tilde{\mathcal{U}}}{_n}\big)\tensor*[^{(1)}]{{\gamma}}{_{-n}^\mu} \nonumber \\[4pt]
&=	{1 \over \sqrt{2\sinh\left(\pi \Omega_{n}\over a\right)}}\left[e^{\pi \Omega_{n} \over 2a}\,	\tensor*[^{(2)}]{\tilde{\gamma}}{_n^\mu} - e^{-{\pi \Omega_{n} \over 2a}}\,\tensor*[^{(1)}]{{\gamma}}{_{-n}^\mu}\right],\\[8pt]
\tensor*[^{\rm F}]{{\beta}}{_n^\mu} &=  \big(\tensor*[^{(1)}]{{\mathcal{W}}}{_n},\tensor*[^{\rm F}]{{\mathcal{U}}}{_n}\big)\tensor*[^{(1)}]{{\gamma}}{_n^\mu} + \big(\tensor*[^{\rm (2)}]{\tilde{\mathcal{W}}}{_n^*},\tensor*[^{\rm F}]{{\mathcal{U}}}{_n}\big)\tensor*[^{\rm (2)}]{\tilde{\gamma}}{_{-n}^\mu} \nonumber \\[4pt]
&=	{1 \over \sqrt{2\sinh\left(\pi \Omega_{n}\over a\right)}}\left[e^{\pi \Omega_{n} \over 2a}\,\tensor*[^{(1)}]{{\gamma}}{_n^\mu} - e^{-{\pi \Omega_{n} \over 2a}}\,\tensor*[^{(2)}]{\tilde{\gamma}}{_{-n}^\mu}\right],\\[8pt]
\tensor*[^{\rm P}]{{\beta}}{_n^\mu} &=  \big(\tensor*[^{\rm (2)}]{{\mathcal{W}}}{_n},\tensor*[^{\rm P}]{{\mathcal{U}}}{_n}\big)\tensor*[^{\rm (2)}]{{\gamma}}{_n^\mu} + \big(\tensor*[^{(1)}]{\tilde{\mathcal{W}}}{_n^*},\tensor*[^{\rm P}]{{\mathcal{U}}}{_n}\big)\tensor*[^{(1)}]{\tilde{\gamma}}{_{-n}^\mu} \nonumber \\[4pt]
&=	{1 \over \sqrt{2\sinh\left(\pi \Omega_{n}\over a\right)}}\left[e^{\pi \Omega_{n} \over 2a}\,\tensor*[^{(2)}]{{\gamma}}{_n^\mu} - e^{-{\pi \Omega_{n} \over 2a}}\,\tensor*[^{(1)}]{\tilde{\gamma}}{_{-n}^\mu}\right].
\end{align}}
Next, for the associated creation operators, we apply the Hermitian conjugate of the above relations, which successively yields
\allowdisplaybreaks{\begin{align}
\tensor*[^{\rm F}]{\tilde{\beta}}{_{-n}^\mu} = \big(\tensor*[^{\rm F}]{\tilde{\beta}}{_n^\mu}\big)^\dagger
&=	{1 \over \sqrt{2\sinh\left(\pi \Omega_{n}\over a\right)}}\left[e^{\pi \Omega_{n} \over 2a}\,\tensor*[^{(1)}]{\tilde{\gamma}}{_{-n}^\mu} - e^{-{\pi \Omega_{n} \over 2a}}\,\tensor*[^{(2)}]{{\gamma}}{_{n}^\mu}\right],\\[6pt]
\tensor*[^{\rm P}]{\tilde{\beta}}{_{-n}^\mu} = \big(\tensor*[^{\rm P}]{\tilde{\beta}}{_n^\mu}\big)^\dagger
&=	{1 \over \sqrt{2\sinh\left(\pi \Omega_{n}\over a\right)}}\left[e^{\pi \Omega_{n} \over 2a}\,\tensor*[^{(2)}]{\tilde{\gamma}}{_{-n}^\mu} - e^{-{\pi \Omega_{n} \over 2a}}\,\tensor*[^{(1)}]{{\gamma}}{_{n}^\mu}\right],\\[6pt]
\tensor*[^{\rm F}]{{\beta}}{_{-n}^\mu} = \big(\tensor*[^{\rm F}]{{\beta}}{_n^\mu}\big)^\dagger
&=	{1 \over \sqrt{2\sinh\left(\pi \Omega_{n}\over a\right)}}\left[e^{\pi \Omega_{n} \over 2a}\,\tensor*[^{(1)}]{{\gamma}}{_{-n}^\mu} - e^{-{\pi \Omega_{n} \over 2a}}\,\tensor*[^{(2)}]{\tilde{\gamma}}{_{n}^\mu}\right],\\[6pt]
\tensor*[^{\rm P}]{{\beta}}{_{-n}^\mu} = \big(\tensor*[^{\rm P}]{{\beta}}{_n^\mu}\big)^\dagger
&=	{1 \over \sqrt{2\sinh\left(\pi \Omega_{n}\over a\right)}}\left[e^{\pi \Omega_{n} \over 2a}\,\tensor*[^{(2)}]{{\gamma}}{_{-n}^\mu} - e^{-{\pi \Omega_{n} \over 2a}}\,\tensor*[^{(1)}]{\tilde{\gamma}}{_{n}^\mu}\right].
\end{align}}
\end{subequations}
Thus, we have derived transformations expressing each right- or left-moving Milne worldsheet operator $\beta_n$ as an inter-mixture of global mode operators $\gamma_n$ (from the same direction) and $\gamma_{-n}$ (from the opposite direction). Furthermore, these global $\gamma$ operators are themselves related to the Minkowski worldsheet oscillators $\alpha$ in a manner determined by the choice of Unruh basis. As a combined effect, it is possible to recombine the relations in \eqref{KBT20} such that they reproduce the Bogoliubov transformations between the Milne worldsheet operators and their inertial counterparts in the Minkowski basis. To realize this, we construct linear combinations of the right- and left-moving Milne worldsheet operators within the same F or P wedge, denoted as $\tilde{\beta}_n^\mu \sim c_1\tensor*[^{\rm F}]{{\beta}}{_n^\mu} + c_2\tensor*[^{\rm F}]{\tilde{\beta}}{_n^\mu}$ and ${\beta}_n^\mu \sim c_1\tensor*[^{\rm P}]{{\beta}}{_n^\mu} + c_2\tensor*[^{\rm P}]{\tilde{\beta}}{_n^\mu}$. By suitably fixing the constants $(c_1, c_2)$, we select a basis in which the inter-mixing of $\big\lbrace \tensor*[^{(1)}]{\tilde{\gamma}}{_n^\mu}, \tensor*[^{(2)}]{\tilde{\gamma}}{_n^\mu}, \tensor*[^{(1)}]{{\gamma}}{_n^\mu}, \tensor*[^{(2)}]{{\gamma}}{_n^\mu}\big\rbrace$ operators aligns exactly with the original Minkowski worldsheet oscillators $\big\lbrace \tilde{\alpha}_n^\mu, {\alpha}_n^\mu \big\rbrace$. This adjustment yields the desired Bogoliubov transformations between the Milne and Minkowski worldsheet operators, valid across the entire spacetime, including both the F and P wedges, as follows:
\begin{subequations}\label{KBT21}
\allowdisplaybreaks{\begin{align}
\tensor*{\tilde{\beta}}{_n^\mu} &=	{1 \over \sqrt{2\sinh\left(\pi \Omega_{n}\over a\right)}}\left[e^{\pi \Omega_{n} \over 2a}\,\tensor*{\tilde{\alpha}}{_n^\mu} - e^{-{\pi \Omega_{n} \over 2a}}\,\tensor*{{\alpha}}{_{-n}^\mu}\right],\\[6pt]
\tensor*{{\beta}}{_n^\mu} &={1 \over \sqrt{2\sinh\left(\pi \Omega_{n}\over a\right)}}\left[e^{\pi \Omega_{n} \over 2a}\,\tensor*{{\alpha}}{_n^\mu} - e^{-{\pi \Omega_{n} \over 2a}}\,\tensor*{\tilde{\alpha}}{_{-n}^\mu}\right],\\[6pt]
\tensor*{\tilde{\beta}}{_{-n}^\mu} &=	{1 \over \sqrt{2\sinh\left(\pi \Omega_{n}\over a\right)}}\left[e^{\pi \Omega_{n} \over 2a}\,\tensor*{\tilde{\alpha}}{_{-n}^\mu} - e^{-{\pi \Omega_{n} \over 2a}}\,\tensor*{{\alpha}}{_n^\mu}\right],\\[6pt]
\tensor*{{\beta}}{_{-n}^\mu} &={1 \over \sqrt{2\sinh\left(\pi \Omega_{n}\over a\right)}}\left[e^{\pi \Omega_{n} \over 2a}\,\tensor*{{\alpha}}{_{-n}^\mu} - e^{-{\pi \Omega_{n} \over 2a}}\,\tensor*{\tilde{\alpha}}{_n^\mu}\right].
\end{align}}	
\end{subequations}  
The above mappings are analogous to the Bogoliubov transformations between Minkowski and Rindler worldsheet operators \cite{Bagchi:2020ats,Bagchi:2021ban}. This resemblance is expected, as the standard time-like entanglement between Milne wedges in quantum field theory is argued to be analogous to the entanglement between space-like Rindler wedges \cite{Socolovsky:2013rga,Olson:2010jy,Higuchi:2017gcd,Quach:2021vzo}. Moreover, both setups employ exactly the same methodology as Unruh \cite{Unruh:1976db}. These novel Bogoliubov transformations, as expressed in \eqref{KBT21}, will be crucial in our exploration of the tensionless limit on the Milne worldsheet from the viewpoint of an observer situated on the Minkowski worldsheet (see \cref{TAW}).

\subsection{The evolving worldsheet vacua}\label{EV}
Now we would like to explore the Milne worldsheet vacuum state $\lvert 0_{\rm K}\rangle$, as defined by the $\beta$ operators in \eqref{KME4}, and establish its mapping with the Minkowski worldsheet vacuum $\lvert 0_{\rm M}\rangle$ of $\alpha$ operators. From the perspective of an observer on the Minkowski worldsheet, the $\beta$ operators are always connected to $\alpha$ operators via the Bogoliubov transformations \eqref{KBT21}. This naturally introduces the notion of a vacuum state $\lvert 0_{\rm K}(a)\rangle$ and associated operators $\big\lbrace \tilde{\beta}_n^\mu(a), {\beta}_n^\mu(a) \big\rbrace$ for the Milne worldsheet that evolve continuously with the time-evolution parameter $a$. This is defined by:
\begin{align}\label{EV1}
\lvert 0_{\rm K}(a)\rangle :~	\tensor*{\tilde{\beta}}{_n^\mu}(a)\lvert 0_{\rm K}(a)\rangle = \tensor*{{\beta}}{_n^\mu}(a)\lvert 0_{\rm K}(a)\rangle  = 0, \enspace \forall~ n>0,
\end{align}
followed by the Bogoliubov transformations (see \cref{KBT21}:
\begin{align}\label{EV2}
\begin{split}
\tensor*{\tilde{\beta}}{_n^\mu}(a) = \beta_+(a)\,\tensor*{\tilde{\alpha}}{_n^\mu} + \beta_-(a)\,\tensor*{{\alpha}}{_{-n}^\mu}, \\[6pt]
\tensor*{{\beta}}{_n^\mu}(a) = \beta_+(a)\,\tensor*{{\alpha}}{_n^\mu} + \beta_-(a)\,\tensor*{\tilde{\alpha}}{_{-n}^\mu}, 
\end{split} 
\end{align} 
where the Bogoliubov coefficients are defined as
\begin{align}\label{EV3}
		\begin{split}
			\beta_+(a) &= {e^{\pi \Omega_{n}\over 2a} \over \sqrt{2\sinh\left(\pi \Omega_{n}\over a\right)}} = {1 \over \sqrt{1 - e^{-{2\pi \Omega_{n} \over a}} }}, \\[10pt]
			\beta_-(a) &= -{e^{-{\pi \Omega_{n}\over 2a}} \over \sqrt{2\sinh\left(\pi \Omega_{n}\over a\right)}} = -{e^{-{\pi \Omega_{n}\over a}} \over \sqrt{1 - e^{-{2\pi \Omega_{n}\over a}} }}.
		\end{split}
\end{align}
We now aim to establish a mapping between the Minkowski worldsheet vacuum $\lvert 0_{\rm M}\rangle$ and the evolving vacuum states $\lvert 0_{\rm K}(a)\rangle$. Following the standard procedure (e.g., see \cite{Smaldone:2018oyl,Katagiri:2023wul,Das:1997gg}), we identify $\beta_+(a) = \cosh\theta_n$ and $\beta_-(a) = \sinh\theta_n$, such that
\begin{align}\label{EV4}
\tanh\theta_n(a) = {\beta_-(a) \over \beta_+(a)} = -e^{-{\pi \Omega_{n}\over a}}.
\end{align}
We then define a generator for the Bogoliubov transformations \eqref{EV2} in the typical form:
\begin{align}\label{EV5}
\mathcal{Q}(\theta_n) = -i \sum_{n=1}^{\infty}\theta_n(a)\left[ \tensor*{\tilde{\alpha}}{_{n}}\cdot\tensor*{{\alpha}}{_{n}}- \tensor*{\tilde{\alpha}}{_{-n}}\cdot\tensor*{{\alpha}}{_{-n}}\right], 
\end{align}
followed by a unitary operator (also recognized as the squeezing operator) given by,
\begin{align}\label{EV6}
U(\theta_n) = e^{i\mathcal{Q}(\theta_n)}.
\end{align}
With the above setup, the relations \eqref{EV3} for the evolving operators can be recast into the Bogoliubov–Valatin transformations:
\begin{align}\label{EV7}
\begin{split}
\tensor*{\tilde{\beta}}{_n^\mu}(a) = U(\theta_n)^\dagger \tensor*{\tilde{\alpha}}{_n^\mu}\, U(\theta_n)  = \cosh\theta_n\,\tensor*{\tilde{\alpha}}{_n^\mu} + \sinh\theta_n\,\tensor*{{\alpha}}{_{-n}^\mu}, \\[6pt]
\tensor*{{\beta}}{_n^\mu}(a)  = U(\theta_n)^\dagger \tensor*{{\alpha}}{_n^\mu}\, U(\theta_n) = \sinh\theta_n\,\tensor*{\tilde{\alpha}}{_{-n}^\mu} + \cosh\theta_n\,\tensor*{{\alpha}}{_n^\mu} .
\end{split} 
\end{align}
These forms provide an alternative expression for the same Milne worldsheet operators as derived in \cref{KBT21}. Consequently, the evolving vacuum states satisfy the following condition for all $n>0$,
\begin{align}\label{EV8}
		\begin{split}
			\tensor*{\tilde{\beta}}{_n^\mu}(a)\lvert 0_{\rm K}(a)\rangle = \left(\tensor*{\tilde{\alpha}}{_n^\mu} + \tanh\theta_n\,\tensor*{{\alpha}}{_{-n}^\mu}\right)\lvert 0_{\rm K}(a)\rangle =  0, \\[6pt]
			\tensor*{{\beta}}{_n^\mu}(a)\lvert 0_{\rm K}(a)\rangle = \left(\tensor*{{\alpha}}{_n^\mu} + \tanh\theta_n\,\tensor*{\tilde{\alpha}}{_{-n}^\mu}\right)\lvert 0_{\rm K}(a)\rangle  = 0.
		\end{split}
\end{align}
The unitary operator $U(\theta_n)$ then connects the evolving vacua $\lvert 0_{\rm K}(a)\rangle$ to the Minkowski worldsheet vacuum $\lvert 0_{\rm M}\rangle$ in the standard manner, which is given by:
\begin{align}\label{EV9}
		\lvert 0_{\rm K}(a)\rangle = U(\theta_n)\lvert 0_{\rm M}\rangle &= \exp\left(-\sum_{n=1}^{\infty}\theta_n(a)\left[ \tensor*{\tilde{\alpha}}{_{-n}}\cdot\tensor*{{\alpha}}{_{-n}} - \tensor*{\tilde{\alpha}}{_{n}}\cdot\tensor*{{\alpha}}{_{n}}\right]\right) \lvert 0_{\rm M}\rangle \nonumber \\[6pt]
		& = \prod_{n=1}^{\infty} {1\over \cosh\theta_n}\exp\left(-{\tanh\theta_n\over n}\,\tensor*{\tilde{\alpha}}{_{-n}}\cdot\tensor*{{\alpha}}{_{-n}}\right)\lvert 0_{\rm M}\rangle,
\end{align} 
where the last step is obtained with the help of standard Gaussian decomposition treatment and the $\alpha$ operator commutation relation \eqref{MW7}. Notably, flipping the positions of the holomorphic and anti-holomorphic oscillators does not alter the typical form of $\lvert 0_{\rm K}(a)\rangle$. Finally, substituting the relations \eqref{EV3} for the $\theta_n$ parameter in terms of $a$, we obtain:
\begin{align}\label{EV10}
		\lvert 0_{\rm K}(a)\rangle = \prod_{n=1}^{\infty} \mathcal{C}_n\exp\left({e^{-{\pi \Omega_{n}\over a}}\over n}\,\tensor*{\tilde{\alpha}}{_{-n}}\cdot\tensor*{{\alpha}}{_{-n}}\right)\lvert 0_{\rm M}\rangle,
\end{align}
where $\mathcal{C}_n = \sqrt{1 - e^{-2\pi \Omega_{n}/a} }$. Therefore, from the perspective of the observer on the Minkowski worldsheet, the new vacua $\lvert 0_{\rm K}(a)\rangle$ of the Milne worldsheet is obtained as a `coherent' or `squeezed' state in terms of the Minkowski worldsheet vacuum $\lvert 0_{\rm M}\rangle$ and its oscillators $\big\lbrace \tilde{\alpha}_n^\mu, {\alpha}_n^\mu \big\rbrace$. This evolving vacua is hence a highly energized state with respect to the inertial one $\lvert 0_{\rm M}\rangle$. Notably, the relations \eqref{EV8} and \eqref{EV9} between $\big\lbrace \tensor*{\tilde{\beta}}{_n^\mu}(a), \tensor*{{\beta}}{_n^\mu}(a) \big\rbrace$ and $\big\lbrace \tilde{\alpha}_n^\mu, {\alpha}_n^\mu \big\rbrace$ are also invertible, yielding:
\begin{align}\label{EV11}
\begin{split}
			\tensor*{\tilde{\alpha}}{_n^\mu}\lvert 0_{\rm M}\rangle = \left(\tensor*{\tilde{\beta}}{_n^\mu}(a) - \tanh\theta_n\,\tensor*{{\beta}}{_{-n}^\mu}(a)\right)\lvert 0_{\rm M}\rangle =  0, \\[6pt]
			\tensor*{{\alpha}}{_n^\mu}\lvert 0_{\rm M}\rangle = \left(\tensor*{{\beta}}{_n^\mu}(a) - \tanh\theta_n\,\tensor*{\tilde{\beta}}{_{-n}^\mu}(a)\right)\lvert 0_{\rm M}\rangle  = 0.
\end{split}
\end{align}
Consequently, we can express $\lvert 0_{\rm M}\rangle$ as a squeezed state of $\lvert 0_{\rm K}(a)\rangle$, given by:
\begin{align}\label{EV12}
		\lvert 0_{\rm M}\rangle = \prod_{n=1}^{\infty} \mathcal{C}_n\exp\left(-{e^{-{\pi \Omega_{n}\over a}}\over n}\,\tensor*{\tilde{\beta}}{_{-n}}(a)\cdot\tensor*{{\beta}}{_{-n}}(a)\right)\lvert 0_{\rm K}(a)\rangle,
\end{align}
which is an inverted perspective of \eqref{EV10}. This is precisely the worldsheet equivalent of the Milne observer perceiving the Minkowski counterpart and its associated vacuum $\lvert 0_{\rm M}\rangle$, which is well-aligned with the timelike Unruh effect discussed in \cref{UE}.

\section{Tensionless limit on Milne worldsheets}\label{TAW}

In this section, we aim to embed the exact notion of ``tensionless-ness'' on the closed string Milne worldsheet evolving with a scaled time-evolution parameter $a$, as depicted in \cref{wkb}. To date, no intrinsic framework exists for tensionless Milne worldsheets, similar to the framework available for the null or Carrollian Minkowski worldsheet in \cref{TMW}. Naively attempting to start from the inertial stage \eqref{IL19} and then applying Unruh's methodology (as employed in \cref{KBT}) for structuring modes on the tensionless Milne worldsheet would likely fail due to the null or degenerate structure of tensionless strings. Therefore, we must consider a limiting approach to the tensile Milne worldsheet setup. A successful example of such an approach has been developed in \cite{Bagchi:2020ats,Bagchi:2021ban} for tensionless Rindler worldsheets. Adapting this idea, we will explore the limit where the Milne worldsheet approaches its horizon, specifically the regime around ${a\over c} \to \infty$. This choice of infinite time-evolution limit on Milne worldsheets is motivated by the role of infinite boost or Carrollian limits in inducing tensionless physics on the Minkowski worldsheet (see \cref{TMW}). Specifically, we are ambitious about conducting a possible litmus test: as the tensile Milne worldsheet approaches its horizon, the Bogoliubov transformations \eqref{KBT21} should truncate in a manner that precisely replicates the mapping \eqref{LA5} between tensionless and tensile worldsheet oscillators.
	
We would like to mention that the set of Bogoliubov transformations \eqref{KBT21} between the $\beta$ and $\alpha$ oscillators of Milne and Minkowski worldsheets are self-consistent in the sense that their respective commutations never disrupt their typical mappings. By reading off the Milne worldsheet parameters $\Omega_{n}$, $\mathcal{L}_{\varphi}$ and $\varphi$, as defined in \cref{wkb}, we impose the limit ${a\over c} \to \infty$ and systematically obtain:
	\begin{align}\label{TAW1}
		{\Omega_{n}\over a} = {2\pi cn \over \ln \left(1 \pm {a\ell\over c}e^{a\psi}\right)} \to {2\pi cn \over \left[\ln \left(\pm {a\ell\over c}\right)+ a\psi\right]} \approx  {2\pi cn\over a\psi}.
	\end{align}       
Evidently, the term on the right-hand side acquires a small value in the near-horizon (or infinite Milne evolution) limit ${a\over c} \to \infty$, causing the transcendental functions within the Bogoliubov coefficients to expand around a small phase factor ${\pi \Omega_{n} \over 2a}$, followed by expressing: 
	\begin{align}\label{TAW2}
		{e^{\pm{\pi \Omega_{n}\over 2a}} \over \sqrt{2\sinh\left(\pi \Omega_{n} \over a\right)}} \approx {1 \over 2}\left(\sqrt{ 2a\over \pi \Omega_{n}} \pm \sqrt{\pi \Omega_{n}\over 2a}\right) + \mathcal{O}\left({\pi\Omega_{n}\over 2a}\right)^{3/2}.
	\end{align}
By implementing these relations, the Bogoliubov transformations \eqref{KBT21} are successively expanded. At leading order (i.e., neglecting terms of $\mathcal{O}\left(\sqrt{c^3 / a^3}\right)$ and higher), we obtain:
	\begin{subequations}\label{TAW3}
		\allowdisplaybreaks{\begin{align} 
				\tensor*[^\infty]{\tilde{\beta}}{_n^\mu} &=	{1 \over 2}\left[\left(\sqrt{\pi^2cn\over a\psi} + \sqrt{a\psi\over \pi^2cn}\right)\tensor*{\tilde{\alpha}}{_n^\mu} + \left(\sqrt{\pi^2cn\over a\psi} - \sqrt{a\psi\over \pi^2cn}\right)\tensor*{{\alpha}}{_{-n}^\mu}\right],\\[10pt]
				\tensor*[^\infty]{\beta}{_n^\mu} &= {1 \over 2}\left[\left(\sqrt{\pi^2cn\over a\psi} + \sqrt{a\psi\over \pi^2cn}\right)\tensor*{{\alpha}}{_n^\mu} + \left(\sqrt{\pi^2cn\over a\psi} - \sqrt{a\psi\over \pi^2cn}\right)\tensor*{\tilde{\alpha}}{_{-n}^\mu}\right],\\[10pt]
				\tensor*[^\infty]{\tilde{\beta}}{_{-n}^\mu} &= {1 \over 2}\left[\left(\sqrt{\pi^2cn\over a\psi} + \sqrt{a\psi\over \pi^2cn}\right)\tensor*{\tilde{\alpha}}{_{-n}^\mu} + \left(\sqrt{\pi^2cn\over a\psi} - \sqrt{a\psi\over \pi^2cn}\right)\tensor*{{\alpha}}{_n^\mu}\right],\\[10pt]
				\tensor*[^\infty]{{\beta}}{_{-n}^\mu} &= {1 \over 2}\left[\left(\sqrt{\pi^2cn\over a\psi} + \sqrt{a\psi\over \pi^2cn}\right)\tensor*{{\alpha}}{_{-n}^\mu} + \left(\sqrt{\pi^2cn\over a\psi} - \sqrt{a\psi\over \pi^2cn}\right)\tensor*{\tilde{\alpha}}{_n^\mu}\right].
		\end{align}}	
	\end{subequations}
Evidently, the above mapping obtained in the near-horizon/light-cone limit of the Milne worldsheet corresponds to the same as the Bogoliubov transformations \eqref{LA5} between tensionless and tensile operators of an intrinsic worldsheet. Comparing \eqref{LA5} and \eqref{TAW3} leads us to the identification of the tensionless state of the Milne worldsheet,
	\begin{align}\label{TAW4}
		\tensor*{\tilde{c}}{_n^\mu} = \tensor*[^\infty]{\tilde{\beta}}{_n^\mu}, \quad \tensor*{{c}}{_n^\mu} = \tensor*[^\infty]{\beta}{_n^\mu},\quad \tensor*{\tilde{c}}{_{-n}^\mu} = \tensor*[^\infty]{\tilde{\beta}}{_{-n}^\mu}, \quad \tensor*{{c}}{_{-n}^\mu} = \tensor*[^\infty]{\beta}{_{-n}^\mu},
	\end{align}      
followed by inferring:
	\begin{align}\label{TAW5}
		\epsilon = {\pi^2n c\over \psi a}, \quad \forall~ n>0.
	\end{align}
Interestingly, the above identifications allow us to estimate the same mode expansion \eqref{IL19} for a Milne worldsheet in the tensionless limit.
	
\subsection{Time evolution and Carrollian limit}\label{EIA}
Now, let us interpret the scenario above, specifically how the concept of `tensionless-ness' enters within the framework of the Milne worldsheet. In other words, how reaching the Milne worldsheet horizons replicates the same physics as a Minkowski worldsheet in the infinite boost limit $\epsilon \to 0$ (refer to \cref{TMW}). By examining the identification in \eqref{TAW5}, we can argue that reaching the tensionless point or the horizon of the Milne worldsheets can be achieved through two possible routes (up to constant factors $n$ and $\psi$):\footnote{Referring to \cref{fig:1}, the $\eta = \text{const.}$ straight lines represent trajectories for a fixed Milne parameter $a$. Route I analytically traces the constant $\xi$ hyperbolas, which degenerate into the light-cone, achieved by increasing $a \to \infty$ while keeping $c$ constant. In contrast, Route II follows the constant $\eta$ straight lines, ultimately collapsing onto the null horizons as $c \to 0$ with $a$ held fixed.}  \begin{align}\label{TAW6}
		\begin{split}
			&\text{Route\, I:} \quad c = \text{constant},\, a \to \infty \implies \epsilon \to 0,\\[8pt] 
			&\text{Route\, II:} \quad a = \text{constant},\, c \to 0 \implies \epsilon \to 0.
		\end{split}
	\end{align}
These two routes to reach the Milne worldsheet light-cone horizon represent distinct physical realizations of the $\epsilon$-parametric evolution, culminating in the infinite boost limit $\epsilon \to 0$, as discussed in \cref{TMW}. Route II, in particular, can be seen as achieving the infinite boost state by directly applying the Carrollian limit \eqref{LA2} on the Milne worldsheet while keeping the time-evolution parameter $a$ constant. In terms of the closed Milne worldsheet coordinates, this can be algebraically implemented by contracting the time coordinate $\xi$ as follows
	\begin{align}\label{TAW7}
		\eta \to \eta, \quad \xi \to \epsilon \xi, \quad \epsilon \to 0.
	\end{align} 
This setup effectively makes $\eta$ very large, allowing one to reach the Milne worldsheet horizon or light-cone. On the other hand, Route I involves an evolution of the Milne worldsheet in parameter $a$, eventually reaching the infinite limit $a \to \infty$. The Bogoliubov transformations of the related evolving oscillators $\big\lbrace \tilde{\beta}_n^\mu(a), {\beta}_n^\mu(a),\tilde{\beta}_{-n}^\mu(a), {\beta}_{-n}^\mu(a) \big\rbrace$ have already been demonstrated in \eqref{EV2}. Consequently, there exists a `flow' of decreasing worldsheet tension throughout the range $a \in [0,\infty]$ as follows
	\begin{align}\label{TAW8}
		\begin{split}
			a = 0 &~~:~ ~ \big\lbrace \tilde{\beta}_n^\mu, {\beta}_n^\mu,\tilde{\beta}_{-n}^\mu, {\beta}_{-n}^\mu \big\rbrace \longrightarrow \big\lbrace \tilde{\alpha}_n^\mu,{\alpha}_n^\mu,\tilde{\alpha}_{-n}^\mu,{\alpha}_{-n}^\mu\big\rbrace, \\[9pt]
			0 < a < \infty &~~:~ ~ \big\lbrace \tilde{\beta}_n^\mu(a), {\beta}_n^\mu(a),\tilde{\beta}_{-n}^\mu(a), {\beta}_{-n}^\mu(a) \big\rbrace, \\[9pt]
			a \to \infty &~~:~ ~\big\lbrace \tilde{\beta}_n^\mu, {\beta}_n^\mu,\tilde{\beta}_{-n}^\mu, {\beta}_{-n}^\mu \big\rbrace \longrightarrow \lbrace \tensor*{\tilde{c}}{_n^\mu},\tensor*{{c}}{_n^\mu},\tensor*{\tilde{c}}{_{-n}^\mu},\tensor*{{c}}{_{-n}^\mu} \big\rbrace.
		\end{split}
	\end{align}
Incorporating this flow into the analysis in \cref{EV}, the tensionless vacuum state $\lvert 0_{c}\rangle$ is obtained as a special limiting case of the relevant evolving vacua $\lvert 0_{\rm K}(a)\rangle$, as defined in \eqref{EV1}, and can be expressed as a highly energized squeezed state in terms of the tensile inertial vacuum $\lvert 0_{\rm M}\rangle$:
	\begin{align}\label{TAW9}
		\lvert 0_{c}\rangle = \lim_{a \to \infty} \lvert 0_{\rm K}(a)\rangle = \mathcal{N}\prod_{n=1}^{\infty}\exp\left({1\over n}\,\tensor*{\tilde{\alpha}}{_{-n}}\cdot\tensor*{{\alpha}}{_{-n}}\right)\lvert 0_{\rm M}\rangle,
	\end{align}
where $\mathcal{N}$ is an (infinite) normalization constant arising after setting $a \to \infty$ in \eqref{EV10}. Notably, the $\epsilon$-parameter evolution also aligns with the framework of \eqref{EV4} and \eqref{EV9}, where $\tanh\theta_n(\epsilon) = {\epsilon - 1 \over \epsilon + 1}$, leading to the same tensionless vacuum in the analogous limit $\epsilon \to 0$. Similarly, using the representation \eqref{EV12}, we can derive the tensile state $\lvert 0_{\rm M}\rangle$ as a coherent state of the tensionless vacuum $\lvert 0_{c}\rangle$, expressed as: 
	\begin{align}\label{TAW10}
		\lvert 0_{\rm M}\rangle  = \hat{\mathcal{N}}\prod_{n=1}^{\infty}\exp\left(-{1\over n}\,\tensor*{\tilde{c}}{_{-n}}\cdot\tensor*{{c}}{_{-n}}\right)\lvert 0_{c}\rangle,
	\end{align}
where $\hat{\mathcal{N}}$ is a different normalization constant in the limit $a \to \infty$. Both mappings \eqref{TAW9} and \eqref{TAW10} suggest a contrasting yet intriguing picture, which we will further explore in the subsequent section.

\section{Discussions}\label{diss}
This section concludes the work by summarizing key findings and discussing their future outlook and potential implications. 

\subsection{Summary of results}\label{summary}
The implications of time-like entanglement within string worldsheet theory remain largely unexplored. In this work, we addressed the crucial question of how a string propagating in a Milne background evolves, with a particular focus on the emergence of a Carrollian limit in the tensionless regime of this time-evolving, non-inertial worldsheet system. Previous studies \cite{Socolovsky:2013rga,Olson:2010jy,Higuchi:2017gcd,Quach:2021vzo} indicate that conventional (R-L) entanglement has a direct counterpart in time-like (F-P) entanglement. This motivated us to develop a Milne worldsheet framework in parallel with the well-established framework of accelerated Rindler worldsheets \cite{Bagchi:2020ats,Bagchi:2021ban}.`

Beginning with the standard setup of a tensile closed-string worldsheet $X^\mu(\tau,\sigma)$ propagating in a $D$-dimensional Minkowski background, we constructed a new target spacetime covering the F-P regions using the transformations \eqref{ES2}, modulated by the Milne parameter $a$. This led to a reparametrization \eqref{CP1} of the worldsheet coordinates $(\tau,\sigma)$ into $(\xi(\tau,\sigma),\eta(\tau,\sigma))$, culminating in the formulation of the Milne worldsheet $\bar{X}^\mu(\xi,\eta)$ with novel features such as folding points and revised periodicity.

We found that the folding points, located at $c\tau = \pm \sigma$, emerge due to the non-inertial nature of the Milne worldsheet, controlled by the characteristic time-evolution parameter $a$. At these points, the induced metric degenerates, and the Milne worldsheet reaches the infinite time-evolution limit ${a\over c} \to \infty$, corresponding to its null horizons. Originally, the periodicity of the worldsheet diverges. However, through a systematic regularization procedure, we derived a new form of periodicity \eqref{CP8} and related mappings \eqref{CP9}, which remained consistent throughout our analysis. The regularization parameter was carefully tuned to coincide with the infinite time-evolution limit of the Milne worldsheet horizons.

Interestingly, the Milne worldsheet exhibits time-like periodicity, in stark contrast to conventional worldsheet theories \cite{Polchinski:1998rq}. We interpreted this by demonstrating that the mapping $X^\mu(\tau,\sigma) \to \bar{X}^\mu(\xi,\eta)$ induces a temporal shift, driven by the Milne parameter $a$, which alters the spatial/temporal nature of the worldsheet and its coordinates. This leads to a rapid shift in periodicity along $\xi$ in the Milne framework, compared to the $\sigma$ direction in its inertial counterpart. In contrast, for a Rindler worldsheet, analogous mappings involve a spatial boost, maintaining the space-like periodicity \cite{Bagchi:2021ban}.

To explore the quantum regime, we performed a quantum mode expansion of the tensile Milne worldsheet, promoting its modes to a global form in terms of the Minkowski basis. This allowed us to link the oscillation operators to their inertial counterparts via Bogoliubov transformations (see \cref{KBT21}). The resulting Bogoliubov coefficients were found to be exponential functions of the Milne parameter $a$, producing a time-evolving vacuum state identified as a squeezed/coherent state relative to the Minkowski vacuum.

Finally, we investigated the transition to the tensionless regime of the Milne worldsheet and identified two limiting routes: (i) taking the infinite time-evolution limit ($a \to \infty$) while keeping the speed of light $c$ fixed, and (ii) sending the speed of light to zero ($c \to 0$) while maintaining a fixed time-evolution frequency $a$. Both routes effectively truncate the tensile Milne framework, collapsing the quantum modes, oscillators, and Bogoliubov transformations into a Carrollian structure—with the first providing an indirect yet non-trivial transition, and the second enforcing a direct Carrollian limit. In both cases, the emergent tensionless regime is governed by Carrollian symmetries, establishing a novel ultra-relativistic limit in the dynamics of time-evolving strings.


\subsection{Outcome and conclusions}\label{conclu}

Building on the detailed construction and quantization of the Milne worldsheet and its tensionless limit, we now draw out the deeper implications of these results for non-inertial string theory. At a conceptual level, this study extends the well-established Rindler worldsheet paradigm \cite{Bagchi:2020ats,Bagchi:2021ban} to a fundamentally distinct yet causally inaccessible domain of non-inertial string dynamics by constructing a time-evolving Milne worldsheet. A key outcome of our analysis is the deep interplay between tensionless string dynamics and emergent Carrollian structures, which becomes apparent upon reaching the null horizons of the Milne worldsheet. This transition necessitates an ultra-high frequency limit in time evolution, setting the Milne framework apart from its Rindler counterpart.

A fundamental distinction in the tensionless limit of the Milne worldsheet, compared to the Rindler worldsheet, lies in the nature of its induced Carrollian structure. While both worldsheets acquire Carrollian properties at their respective infinite limits, their underlying mechanisms and geometric interpretations diverge significantly. The Rindler limit is dictated by an infinite boost, producing an emergent Carrollian symmetry along spatial directions and forming a Rindler horizon endowed with a Carrollian structure, where entanglement between the left (L) and right (R) Rindler wedges encodes the tensionless structure. In contrast, the Milne worldsheet undergoes an infinite rescaling of its temporal coordinate, leading to a Milne horizon with Carrollian structure, governed by time-like entanglement between past (P) and future (F) wedges. Another key distinction is that while the Rindler limit preserves space-like periodicity, the Milne limit enforces an inherently time-like periodicity, reshaping the structure of worldsheet evolution.

Consequently, the Milne framework broadens the Carrollian analysis of tensionless strings to both dynamical and static non-inertial settings, where the frequency of time evolution, rather than acceleration, governs the dynamics. This reveals a striking correspondence: in non-inertial string theory, acceleration and time-evolution frequency serve as structurally analogous parameters. As a result, an observer on the Milne worldsheet can undergo non-inertial evolution toward the Carrollian point even while remaining static—a scenario impossible for an observer on a Rindler worldsheet, who must be accelerating to reach the same final state. These structural differences introduce an intrinsic rigidity that prevents a direct mapping between the two frameworks, underscoring the novelty of this work.

Our findings decisively establish that non-inertial string dynamics emerge in two structurally distinct yet complementary ways—governed either by worldsheet acceleration (Rindler) or the frequency of its time evolution (Milne). Even more remarkably, we uncover a duality-like correspondence between Rindler and Milne worldsheets near their respective null horizons, both exhibiting identical tensionless physics. This strongly suggests the existence of a novel time-like counterpart to the ultra-relativistic (Carrollian) or infinite-boost limit, offering a fresh perspective on non-inertial string theory and expanding its conceptual framework. Furthermore, while Rindler and Milne worldsheets attain their tensionless limits through distinct physical mechanisms, they ultimately converge to an equivalent emergent Carrollian structure. Most strikingly, despite their fundamentally different evolutions, the respective non-inertial string vacua collapse into identical open string bound states at the tensionless limit, reinforcing the universality of the phenomenon.

These outcomes establish a key conclusion: time-like (F-P) entanglement in Milne worldsheets is not merely analogous to but fundamentally equivalent to standard (R-L) entanglement in Rindler worldsheets, both occurring in distinct pairs of causally disconnected regions of the background spacetime. This correspondence provides a robust ``time-like window'' for probing the tensionless regime of fundamental strings. Thus, we have uncovered a novel route to reaching the null horizons of non-inertial worldsheets, governed by an equivalent degenerate metric structure. 

More fundamentally, the complete characterization of non-inertial string physics follows a 1-to-2 correspondence between the observer on the Minkowski worldsheet and the pair of observers on the Milne and Rindler worldsheets. This arises from the fact that the Minkowski worldsheet’s target spacetime is precisely partitioned—half into the F and P regions and half into the R and L regions—between the Milne and Rindler worldsheets, thereby structuring their respective target backgrounds. In other words, the physics of non-inertial strings, as perceived from the Minkowski worldsheet, emerges from an equal reconstruction of information from both the Milne and Rindler counterparts. In this regard, the present work establishes that the Milne branch of non-inertial worldsheets is not merely a complementary extension but an indispensable missing component for achieving a more unified perspective on non-inertial string dynamics. Crucially, this scenario undergoes a fundamental transformation upon reaching their Carrollian points via their respective limiting processes. It is only at this stage that the two worldsheets effectively align along a common null line or regime, despite originating from causally distinct routes, ultimately converging onto an equivalent physical description. More precisely, the time-like Carrollian limit reshapes the 1-to-2 correspondence into a 1-to-1 unification, wherein the Minkowski worldsheet observer effectively perceives the Milne and Rindler worldsheet observers as gaining mutual access to each other’s tensionless physics—an interaction that was previously forbidden in the entire tensile regime. This unification bridges the gap between acceleration-driven and time-evolution-driven worldsheets at their tensionless limit.

Given these insights, we anticipate that the Milne worldsheet framework will play a pivotal role in unveiling the fundamental significance of Carrollian structures in organizing the quantum properties of tensionless strings. The emergence of a Carrollian limit in this formulation not only provides a new approach to the tensionless regime but also suggests deeper connections to the causal structure of spacetime, reinforcing the role of non-inertial evolution in string theory. This framework naturally extends the well-established paradigm of null Rindler worldsheets to a structurally distinct yet analogous setting, where time evolution, rather than acceleration, governs non-inertial dynamics. As a result, our findings offer a new perspective on the interplay between string theory, quantum field theory, and the geometric aspects of Carrollian physics.

Taken together, these observations suggest that the Milne construction goes beyond the conventional paradigm by revealing an uncharted route toward the ultra-relativistic sector of string theory through ultra-high-frequency time evolution. So far, the tensionless or Carrollian sector has predominantly been realized either through worldsheet contractions associated with infinite boosts in inertial settings~\cite{Schild:1976vq,Isberg:1993av,Bagchi:2013bga,Bagchi:2015nca}, or through the infinite-acceleration limit of Rindler worldsheets~\cite{Bagchi:2020ats,Bagchi:2021ban}, where the Carrollian regime emerges as a manifestation of the underlying boost structure. In contrast, the present construction reveals that ultra-high-frequency time evolution itself dynamically drives the worldsheet toward an intrinsically ultra-relativistic regime without requiring an underlying infinite-boost interpretation. In this sense, the ultra-high-frequency limit provides a natural realization of Carrollian worldsheet contractions~\cite{SenGupta:1966qer,Bagchi:2013bga,Bagchi:2015nca}, where the temporal direction becomes effectively compressed while the spatial worldsheet direction remains absolute. In fact, such a timelike realization already appeared in the original formulation of Carrollian structures~\cite{SenGupta:1966qer}, which now finds a concrete realization at the worldsheet level.

From this perspective, the ultra-high-frequency time-evolution limit on the worldsheet acquires important implications for the tensionless and non-Lorentzian sectors of string theory through a genuinely distinct timelike realization of the worldsheet symmetry algebras. In the present setting, the disconnected future-past sectors of the worldsheet may be viewed as carrying the two-copy Virasoro structures, rather than the conventional left-right decomposition of Rindler worldsheets. As the ultra-high-frequency limit is approached, these disconnected sectors dynamically compress toward a common null-horizon structure, marking the emergence of a single Carrollian or BMS-type algebra from the underlying two-copy Virasoro symmetry. Therefore, the ultra-high-frequency time-evolution limit of the worldsheet Virasoro algebra and the conventional infinite acceleration or boost limit lead to the same BMS symmetry structure. The emergence of the BMS algebra from the Virasoro symmetry in this way suggests that the distinction between the two realizations of ultra-relativistic limits becomes effectively blurred near the null horizons. This indistinguishability could become particularly relevant for understanding strings approaching black-hole horizons from interior and exterior regions, where both infinite time-evolution and infinite-acceleration limits may naturally lead to the same tensionless Carrollian phase.

\subsection{Further discussions and implications}\label{fdiss}
In this note, we extend the discussion by thoroughly analyzing the critical implications of tensionless physics and exploring the limiting processes that shape the Milne worldsheet framework. These investigations are essential, as they provide a fundamental litmus test for establishing the Carrollian and tensionless nature of the time-evolving Milne worldsheet formulated in this work.

\subsubsection{Emergence of open string states}\label{OSS}
In string theory, open strings are often viewed as defects on the boundaries of closed string worldsheets. The fundamental boundary conditions for open strings can be expressed in terms of closed string boundary states as follows \cite{Blumenhagen:2009zz}  
	\begin{align}\label{TAW11}
		\begin{split}
			\partial_\tau X^\mu (\sigma,\tau)\big\rvert_{\tau =0} \lvert\mathfrak{B}_{\rm N}\rangle = 0 & ~\qquad  \text{Neumann condition}, \\[8pt]
			\partial_\sigma X^\mu (\sigma,\tau)\big\rvert_{\tau =0} \lvert \mathfrak{B}_{\rm D}\rangle = 0 & ~\qquad  \text{Dirichlet condition},
		\end{split}
	\end{align}
where $\lvert\mathfrak{B}_{\rm N}\rangle$ and $\lvert\mathfrak{B}_{\rm D}\rangle$ are regarded as Neumann and Dirichlet boundary states, respectively. Using the Minkowski worldsheet mode expansion \eqref{MW10}, the boundary conditions \eqref{TAW11} can be expressed in terms of the inertial $\alpha$-oscillators as
	\begin{align}\label{TAW12}
		\begin{split}
			\left(\tilde{\alpha}_n^\mu + {\alpha}_{-n}^\mu\right)\lvert\mathfrak{B}_{\rm N}\rangle = \left({\alpha}_n^\mu + \tilde{\alpha}_{-n}^\mu\right)\lvert\mathfrak{B}_{\rm N}\rangle =0, \\[6pt]
			\left(\tilde{\alpha}_n^\mu - {\alpha}_{-n}^\mu\right)\lvert\mathfrak{B}_{\rm D}\rangle = \left({\alpha}_n^\mu - \tilde{\alpha}_{-n}^\mu\right)\lvert\mathfrak{B}_{\rm D}\rangle =0,
		\end{split}
	\end{align} 
for each $n > 0$. These conditions, which relate the right- and left-moving closed string modes acting on the boundary states, are the so-called \textit{gluing conditions}. In other words, the conditions \eqref{TAW12} effectively treat the two halves of the string as a pair of open strings by gluing them together. The conditions \eqref{TAW12} can be solved explicitly by following the procedure used to obtain the squeezed state \eqref{EV9} respectively for $\tanh \theta_n = \pm 1$, yielding:     
	\begin{align}\label{TAW13}
		\begin{split}
			\lvert\mathfrak{B}_{\rm N}\rangle &= \mathcal{N}_{\rm N}\prod_{n=1}^{\infty}\exp\left(-{1\over n}\,\tensor*{\tilde{\alpha}}{_{-n}}\cdot\tensor*{{\alpha}}{_{-n}}\right)\lvert 0_{\rm M}\rangle, \\[6pt]
			\lvert\mathfrak{B}_{\rm D}\rangle &= \mathcal{N}_{\rm D}\prod_{n=1}^{\infty}\exp\left({1\over n}\,\tensor*{\tilde{\alpha}}{_{-n}}\cdot\tensor*{{\alpha}}{_{-n}}\right)\lvert 0_{\rm M}\rangle.
		\end{split}
	\end{align}
Interestingly, the tensionless worldsheet vacuum state representations in \eqref{TAW9} and \eqref{TAW10} can be identified as the typical boundary states depicted in \eqref{TAW13}. If we elaborate further, it becomes evident that the coherent representation \eqref{TAW9} of the tensionless Milne vacuum $\lvert 0_{c}\rangle$, expressed in terms of the usual inertial worldsheet vacuum $\lvert 0_{\rm M}\rangle$ and its $\alpha$-oscillators, forms a Dirichlet boundary state in all spacetime directions. On the other hand, in terms of the $\lvert 0_{c}\rangle$ vacuum and its $c$-oscillators, the coherent state form \eqref{TAW10} of the inertial vacuum $\lvert 0_{\rm M}\rangle$ corresponds to a Neumann boundary state. This observation indicates that, as the tension vanishes, closed-string worldsheets transition to an open-string description. To validate this phenomenon, we can verify whether these coherent vacuum states $\lvert 0_{c}\rangle$ and $\lvert 0_{\rm M}\rangle$ satisfy the gluing conditions in a form similar to \eqref{TAW12}, thereby confirming their identification as worldsheet boundary states. In terms of the inertial $\alpha$-operators, the evolving Milne vacua $\lvert 0_{\rm K}(a)\rangle$ (as defined in \eqref{EV8}) can be expressed as
	\begin{align}\label{TAW14}
		\left(\tensor*{\tilde{\alpha}}{_n^\mu} -e^{-{\pi \Omega_{n}\over a}}\tensor*{{\alpha}}{_{-n}^\mu}\right)\lvert 0_{\rm K}(a)\rangle = \left(\tensor*{{\alpha}}{_n^\mu} -e^{-{\pi \Omega_{n}\over a}}\tensor*{\tilde{\alpha}}{_{-n}^\mu}\right)\lvert 0_{\rm K}(a)\rangle = 0.
	\end{align}   
Similarly, with respect to the evolving Milne operators $\beta(a)$, the Minkowski vacuum $\lvert 0_{\rm M}\rangle$ (as defined in \eqref{EV11}) reads,   
	\begin{align}\label{TAW15}
		\left(\tensor*{\tilde{\beta}}{_n^\mu}(a) + e^{-{\pi \Omega_{n}\over a}}\tensor*{{\beta}}{_{-n}^\mu}(a)\right)\lvert 0_{\rm M}\rangle = \left(\tensor*{{\beta}}{_n^\mu}(a) + e^{-{\pi \Omega_{n}\over a}}\tensor*{\tilde{\beta}}{_{-n}^\mu}(a)\right)\lvert 0_{\rm M}\rangle = 0.
	\end{align} 
If we now take the typical tensionless limit ${a} \to \infty$ (keeping $c$ constant), the evolving Milne oscillators and associated vacua translate into tensionless $c$-operators and the vacuum state $\lvert 0_{c}\rangle$ in \eqref{TAW14} and \eqref{TAW15}, yielding:
	\begin{align}\label{TAW16}
		\begin{split}
			\left(\tilde{\alpha}_n^\mu - {\alpha}_{-n}^\mu\right)\lvert 0_{c}\rangle &= \left({\alpha}_n^\mu - \tilde{\alpha}_{-n}^\mu\right)\lvert 0_{c}\rangle =0, \\[6pt]
			\left(\tilde{c}_n^\mu + {c}_{-n}^\mu\right)\lvert 0_{\rm M}\rangle &= \left({c}_n^\mu + \tilde{c}_{-n}^\mu\right)\lvert 0_{\rm M}\rangle =0.
		\end{split}
	\end{align} 
These are clearly the gluing conditions that identify the tensionless Milne vacuum $\lvert 0_{c}\rangle$ and the tensile Minkowski vacuum $\lvert 0_{\rm M}\rangle$ as Dirichlet and Neumann boundary states, respectively. Thus, taking the tensionless limit on any closed string worldsheet results in the emergence of open strings. This situation, where defect states appear on the closed worldsheets, arises when all points collapse onto the null or horizon lines $c\tau = \pm \sigma$. To summarize, from the perspective of an observer in the inertial worldsheet vacuum $\lvert 0_{\rm M}\rangle$, the evolving Milne vacuum at the tensionless limit $\lvert 0_{c}\rangle$ is interpreted as a Dirichlet boundary state, representing a spacetime point known as a D-instanton \cite{Polchinski:1998rq}. Conversely, from the viewpoint of the null Milne vacuum $\lvert 0_{c}\rangle$, the Minkowski vacuum $\lvert 0_{\rm M}\rangle$ serves as a Neumann boundary state, corresponding to a space-filling D-brane \cite{Bagchi:2019cay}.
\begin{figure}[t!]
	\centering
	\includegraphics[width=1.05\textwidth]{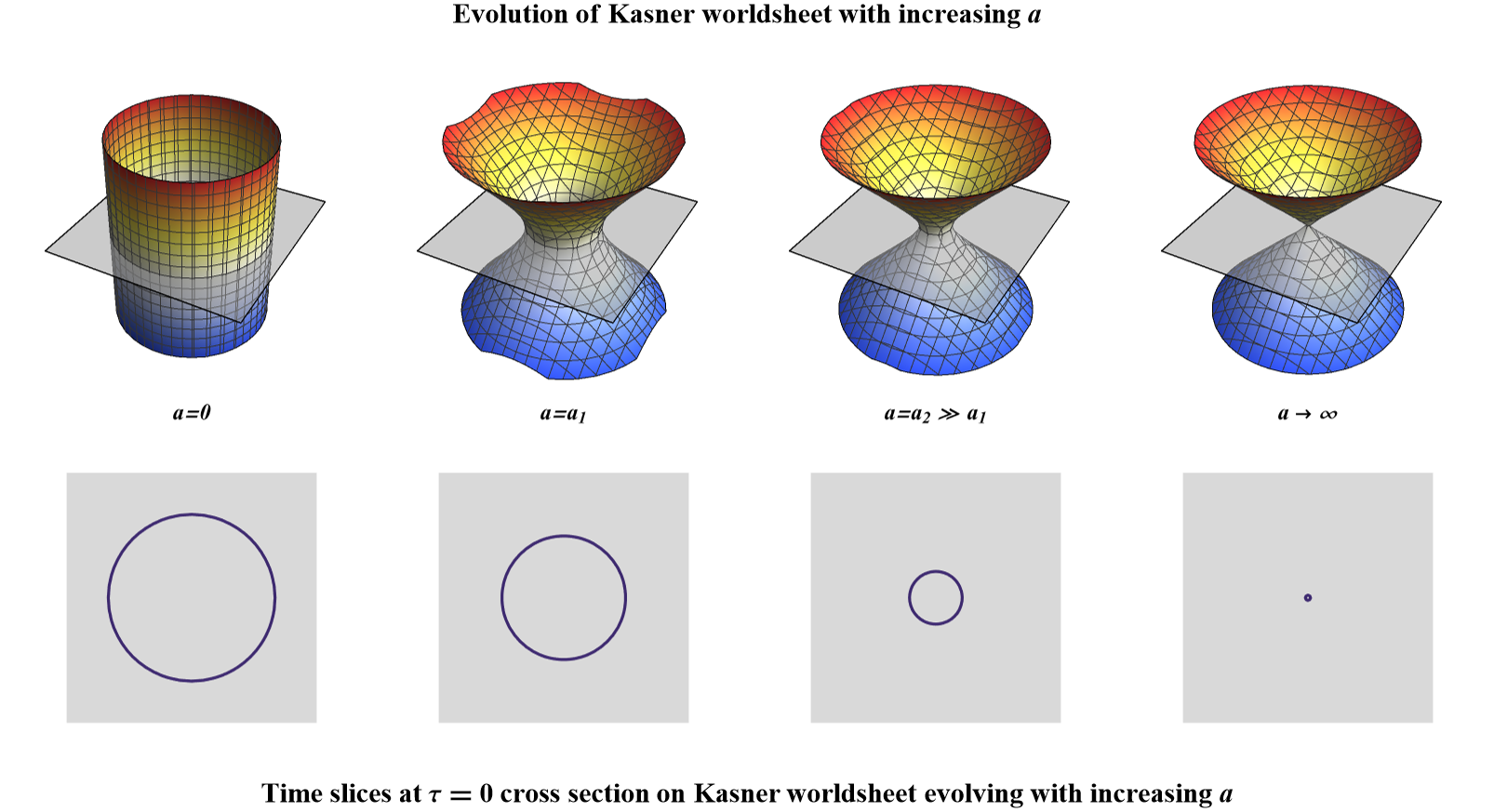}
	\caption{Illustration of a closed string Milne worldsheet deforming with increasing $a$ as seen from the Minkowski worldsheet. In the upper half, the cylindrical snapshot at $a=0$ shows the inertial state. As $a$ increases ($a_2 \gg a_1$) keeping $c$ constant, the worldsheet distorts into hyperboloids with growing eccentricity, eventually transforming into a light-cone at $a \to \infty$. The lower half shows the $\tau = 0$ cross-sectional snapshots of this evolution.}\label{fig:2}
\end{figure}
\begin{figure}[t!]
	\centering
	\includegraphics[width=1.01\textwidth]{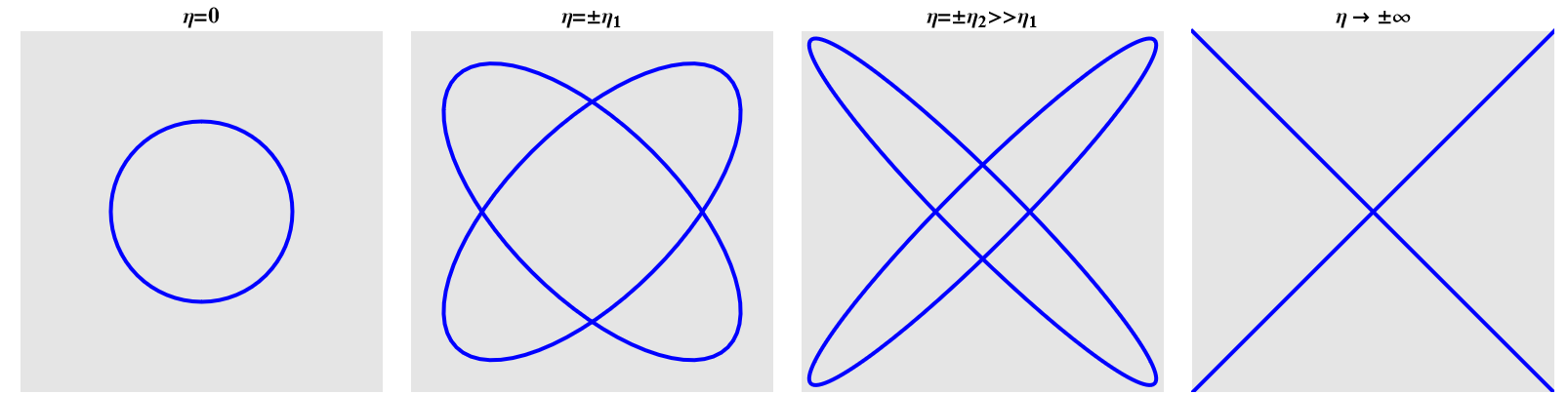}
	\caption{Illustration of a closed string on the Minkowski worldsheet deforming as observed from the Milne worldsheet along changing constant $\eta$ lines, as shown in \cref{fig:1}. The circular snapshot on the left represents the pure tensile state. As the slope of the $\eta$ lines decreases or increases ($\eta_2 \gg \eta_1$), the string distorts into an ellipse, ultimately aligning with the light-cone at $\eta \to \pm\infty$.}\label{fig:3}
\end{figure}

The above scenario of open string physics emerging from closed strings directly corresponds to the so-called \textit{null string complementarity}, which has already been observed for a Rindler worldsheet \cite{Bagchi:2020ats,Bagchi:2021ban}. Similarly, for the current case of the Milne worldsheet, this idea can be physically interpreted as follows. From the perspective of an observer on the tensile Minkowski worldsheet, the Milne worldsheet with increasing $a$ evolves from a cylinder to distorted hyperboloids (which are a natural generalizations of Milne particle worldlines), finally transforming into a light-cone in the tensionless limit $a \to \infty$. For the Dirichlet boundary state \eqref{TAW9}, when zooming in on the $\tau = 0$ cross-section, where the boundary states \eqref{TAW12} are defined, the evolution appears as a circle with decreasing radius, eventually collapsing to a point, representing the emergence of a D-instanton (see \cref{fig:2}). Conversely, an observer on the Milne worldsheet, as $a \to \infty$, will perceive the Minkowski worldsheet as becoming increasingly elongated as its tension decreases, ultimately becoming infinitely stretchable. This interpretation views the Neumann boundary state \eqref{TAW10} as D-branes progressively filling all of spacetime. This latter viewpoint physically manifests for different observers on the Milne worldsheet, positioned along constant $\eta$ lines from $\eta = 0$ to $\eta \to \pm \infty$ along a fixed $\xi$ hyperbola (e.g., see \cref{fig:3}), as they witness the complete sequence of deformation from a tensile to a tensionless Minkowski worldsheet. Thus, as the closed string worldsheet becomes null or tensionless, both Milne and Minkowski observers gain complementary perspectives on the emergence of open string physics. This entire phenomenon associated with the Milne worldsheet in the Carrollian limit serves as a strong `time-like window' into null strings. For consistency, readers may compare this with the analogous observation by Bagchi et al. for the case of accelerated Rindler worldsheet in \cite{Bagchi:2020ats} (see, for instance, Fig. 2, Fig. 3, and the related discussions). Evidently, both approaches to studying tensionless strings are complementary, ultimately leading to equivalent physical insights and outcomes.

\subsubsection{Time-like Unruh effect on tensionless Milne worldsheets}\label{UE}
The observer on the Milne worldsheet will see particles defined by the $\beta$ operators in the Minkowski worldsheet vacuum $\lvert 0_{\rm M}\rangle$. In this context, we can define the following number operators for the $\beta$ particles as
{\allowdisplaybreaks
\begin{align}\label{UE1}
\tensor*{\tilde{N}}{_{n}} = {1 \over n}\tensor*{\tilde{\beta}}{_{-n}}\cdot\tensor*{\tilde{\beta}}{_{n}} =\left\{
\begin{array}{lc}\nonumber
{1 \over n}\tensor*[^{\rm F}]{\tilde{\beta}}{_{-n}}\cdot\tensor*[^{\rm F}]{\tilde{\beta}}{_{n}} = \tensor*[^{\rm F}]{\tilde{N}}{_{n}} & ~: \ \rm F \\~\\
{1 \over n}\tensor*[^{\rm P}]{\tilde{\beta}}{_{-n}}\cdot\tensor*[^{\rm P}]{\tilde{\beta}}{_{n}} = \tensor*[^{\rm P}]{\tilde{N}}{_{n}} & ~: \ \rm P
\end{array} \right.
\\\nonumber~\\
\tensor*{{N}}{_{n}}  = {1 \over n}\tensor*{{\beta}}{_{-n}}\cdot\tensor*{{\beta}}{_{n}} =\left\{
\begin{array}{lc}
{1 \over n}\tensor*[^{\rm F}]{{\beta}}{_{-n}}\cdot\tensor*[^{\rm F}]{{\beta}}{_{n}} = \tensor*[^{\rm F}]{{N}}{_{n}} & ~: \ \rm F \\~\\
{1 \over n}\tensor*[^{\rm P}]{{\beta}}{_{-n}}\cdot\tensor*[^{\rm P}]{{\beta}}{_{n}} = \tensor*[^{\rm P}]{{N}}{_{n}} & ~: \ \rm P
\end{array} \right.
\end{align}}
where all spacetime indices are contracted. The expectation values of the above-listed number operators in state $\lvert 0_{\rm M}\rangle$ describe the average number of particles detected by the Milne worldsheet observer in the Minkowski worldsheet vacuum. For an example, the expected number of particles of frequency $\Omega_{n}$ associated with $\tensor*{\tilde{N}}{_{n}}$ operator is given by,
	\begin{align}\label{UE2}
		\langle0_{\rm M}\lvert\tensor*{\tilde{N}}{_{n}}\rvert0_{\rm M}\rangle &= \sum_{\Lambda,\Lambda^\prime = \rm F,P} {1 \over {2\sinh\left(\pi \Omega_{n}\over a\right)}}\langle0_{\rm M}\left| {1 \over n} e^{-{\pi \Omega_{n}\over a}}\, \tensor*[^{\rm \Lambda}]{\tilde{\gamma}}{_{-n}}\cdot\tensor*[^{\rm {\Lambda^\prime}}]{\tilde{\gamma}}{_{n}} \right| 0_{\rm M}\rangle \nonumber \\[6pt]
		&= {e^{-{\pi \Omega_{n}\over a}} \over {2\sinh\left(\pi \Omega_{n}\over a\right)}}\, \delta_{0,0}= {{1} \over {e^{\Omega_{n}\over a/2\pi}-1}}\, \delta_{0,0}.
	\end{align}
Similarly, for the case of $\tensor*{{N}}{_{n}}$ operator, we can have
	\begin{align}\label{UE3}
		\langle0_{\rm M}\lvert\tensor*{{N}}{_{n}}\rvert0_{\rm M}\rangle = {{1} \over {e^{\Omega_{n}\over a/2\pi}-1}}\, \delta_{0,0}.
	\end{align} 
Note that the infinite factor $\delta_{0,0}$ both in \eqref{UE2} and \eqref{UE3} arises from the commutator of $\gamma$ operators,
	\begin{align}\label{UE4}
		\big[\tensor*[^{\rm \Lambda}]{\tilde{\gamma}}{_n^\mu},\tensor*[^{\rm {\Lambda^\prime}}]{\tilde{\gamma}}{_m^\nu}\big] = \big[\tensor*[^{\rm \Lambda}]{\gamma}{_n^\mu},\tensor*[^{\rm {\Lambda^\prime}}]{\gamma}{_m^\nu}\big] = n\eta^{\mu\nu}\delta_{\Lambda,\Lambda^\prime}\delta_{n+m,0}, \qquad \forall~ \Lambda,\Lambda^\prime = \rm (F,P)
	\end{align}
It is important to notice that the right-hand side of \eqref{UE2} and \eqref{UE3} precisely represents a spectrum of thermal radiation (i.e., Bose-Einstein distribution) with $\hbar = k_{\rm B} = 1$  and an absolute temperature,
	\begin{align}\label{UE5}
		\tensor*{{T}}{_{\rm U}} = {a \over 2\pi}. 
	\end{align}
This phenomenon corresponds to the well-known timelike Unruh effect \cite{Olson:2010jy}. Thus, an observer on the Milne worldsheet evolving with the time-evolution parameter $a$ will detect particles in the vacuum state of the Minkowski worldsheet at the Unruh temperature given by \eqref{UE5}, while an observer on the Minkowski worldsheet will detect no particles in its own vacuum state. In this context, we note that the intrinsic nature of the time-evolution parameter $a$ (as compared to the Rindler acceleration) makes the timelike Unruh temperature \eqref{UE5} more feasible for detection \cite{Olson:2010jy}, revealing a novel aspect of time-like entanglement as a probe for tensionless non-inertial worldsheets.
	
\subsubsection{Triggering Hagedorn physics through time-like Unruh temperature}\label{HU}
It is well understood that the framework of tensionless strings is intimately connected to string theory at the Hagedorn temperature $T_{\rm H}$. Arguably, $T_{\rm H}$ defines an extreme and absolute limiting temperature within the string theory framework, at which a novel phase transition occurs, introducing new degrees of freedom \cite{Atick:1988si}. Beyond this point, the canonical partition function describing all single-particle states within the string framework diverges, and the strings effectively become tensionless \cite{Pisarski:1982cn, Olesen:1985ej}. For a string system continuously evolving with temperature $T_{\rm S}$, the effective tension $\mathcal{T}_{\rm eff}$ is related to the Hagedorn temperature $T_{\rm H}$ via:
	\begin{align}\label{HU1}
		\mathcal{T}_{\rm eff} = \mathcal{T}\sqrt{1 - \left({T_{\rm S} \over T_{\rm H}}\right)^2},
	\end{align} 
where $\mathcal{T}$ is the intrinsic string tension as defined in \eqref{MW2}. Clearly, $\mathcal{T}_{\rm eff} \approx \mathcal{T}$ as $T_{\rm S} \to 0$, and $\mathcal{T}_{\rm eff} \to 0$ as the system temperature approaches $T_{\rm S} \to T_{\rm H}$. In this context, the analysis of tensionless worldsheets in \cref{TMW,TAW} is found to be highly relevant and useful. More precisely, the tensionless vacuum state $\lvert 0_{c}\rangle$ (which is entirely distinct from the tensile vacuum states $\lvert 0_{\rm M}\rangle$ and $\lvert 0_{\rm K}\rangle$) is conjectured to be the worldsheet manifestation of the emergent long string as the system approaches the $T_{\rm H}$ limit \cite{Bagchi:2015nca}. This conjecture is supported by the observation that $\lvert 0_{c}\rangle$ can be obtained as a squeezed or coherent state \eqref{TAW9} of the tensile worldsheet theory, representing a highly energized state analogous to the string system near the Hagedorn temperature. The new degrees of freedom induced by the Hagedorn phase transition in string theory can be interpreted as excitations of the worldsheet vacuum $\lvert 0_{c}\rangle$ through the action of its creation operators $\lbrace \tensor*{\tilde{c}}{_{-n}^\mu},\tensor*{{c}}{_{-n}^\mu}\rbrace$. 
	
We now aim to explore how the timelike Unruh physics induced on the inertial worldsheet from the perspective of the Milne observer is connected to Hagedorn physics. Specifically, we investigate how the timelike Unruh temperature $T_{\rm U}$ (as derived in \eqref{UE5}) at the worldsheet tensionless limit triggers the same physics as that emerging from the Hagedorn phase transition at $T_{\rm S} \to T_{\rm H}$. In this analysis, we interpret the worldsheet evolving toward its tensionless point in terms of $\epsilon$-parameter flow approaching the $\epsilon \to 0$ limit. Additionally, we found that the $\epsilon$-parameter flow can be manifested in terms of $a$-parameter evolution, which is mapped via:  
	\begin{align}\label{HU2}
		\tanh\theta  = {\epsilon - 1 \over \epsilon + 1} = -e^{-{\pi \Omega_{n}\over a}},
	\end{align}
where $\tanh\theta$ represents the conventional ratio of Bogoliubov coefficients from the transformations \eqref{EV2} and \eqref{LA5}. From this stage, it is straightforward to translate everything into the language of a worldsheet evolving with increasing timelike Unruh temperature $T_{\rm U}$. In this process, we observe that the Minkowski worldsheet follows a thermodynamic relation analogous to \eqref{HU1} as seen by the Milne worldsheet observer, yielding:       
	\begin{align}\label{HU3}
		\mathcal{T}_{\rm eff} = \mathcal{T}\epsilon, \quad \epsilon = \tanh\left({\Omega_{n}\over 4T_{\rm U}}\right).	
	\end{align}
By comparing the typical effective tension forms \eqref{HU1} and \eqref{HU3}, we identify
	\begin{align}\label{HU4}
		T_{\rm S} =  T_{\rm H}\sqrt{1 - \tanh^2\left({\Omega_{n}\over 4T_{\rm U}}\right)}.
	\end{align}
Thus, the observer on the Milne worldsheet will see that as the timelike Unruh temperature approaches infinity, the Minkowski worldsheet vacuum undergoes a transition from $\lvert 0_{\rm M}\rangle$ to $\lvert 0_{c}\rangle$ at $\epsilon \to 0$, mirroring the Hagedorn phase transition $T_{\rm S} \to T_{\rm H}$ of tensionless strings, i.e.,     
	\begin{align}\label{HU5}
		T_{\rm U} \to \infty \implies  \epsilon \to 0, \quad T_{\rm S} \approx T_{\rm H}.
	\end{align}
The above phenomenon of manifesting the Hagedorn phase transition on worldsheets suggests a different form of the \textit{null string complementarity} discussed in \cref{OSS}. This idea can be explained as follows. From the perspective of an observer on the Minkowski worldsheet, the tensionless vacuum $\lvert 0_{c}\rangle$ is arguably experiencing Bose-Einstein condensation (BEC) as a consequence of the Hagedorn phase transition \cite{Bagchi:2019cay}.\footnote{Readers may wonder how BEC, which typically occurs at extremely low (or absolute) temperatures, corresponds to the Hagedorn phase transition, a high-energy phenomenon. To clarify, note that for fundamental strings, the Hagedorn temperature is given by $T_{\rm H} = {1 \over 2\pi\sqrt{2\alpha^\prime}}$. Therefore, as the worldsheet approaches the Hagedorn phase transition by dialing the tensionless limit $\alpha^\prime \to \infty$, this triggers a very low value for $T_{\rm H}$, thereby connecting the worldsheet vacuum with BEC.} In contrast, an observer on the Milne worldsheet would observe $\lvert 0_{c}\rangle$ undergoing the same Hagedorn phase transition at a high temperature $T_{\rm U} \to \infty$, resulting in the worldsheet vacuum bubbling with high-energy particles uniformly spread across $\lvert 0_{c}\rangle$. This scenario is complementary to one where BEC formation is suppressed. This complementarity picture reflects the same ambiguity in interpreting the tensionless worldsheet vacuum as both a spacetime point (i.e., space-filling D-instanton) and an infinitely long string filling the entire spacetime (i.e., space-filling D-branes). 

Notably, the BEC interpretation, open-string emergence, null string complementarity, and Hagedorn physics together constitute hallmark signatures of the Carrollian and tensionless sectors of string theory and therefore provide powerful consistency checks for validating the tensionless nature of a given worldsheet construction. In this regard, it is noteworthy that the same interconnected structure first appeared for the null Minkowski worldsheet \cite{Bagchi:2019cay}, was subsequently identified for accelerated Rindler worldsheets \cite{Bagchi:2020ats,Bagchi:2021ban}, and now emerges naturally within the Milne construction presented in this work. The recurrence of this common pattern across these distinct settings further supports the universality of the underlying tensionless-string physics and highlights the closely analogous physical interpretations associated with accelerated and time-evolving worldsheets in their respective tensionless limits.



\subsection{Future directions and questions}\label{outlook}

Given that the preceding sections have successfully addressed the key questions posed in \cref{intro}, the findings of this paper might provide a foundation for deeper investigations into non-inertial tensionless string dynamics, particularly in relation to holography, quantum gravity, and early universe cosmology. A promising avenue for future study is the behavior of time-evolving strings in the presence of black hole horizons, where Carrollian structures may play a crucial role in near-horizon physics and the black hole information paradox. Another crucial aspect is the impact of the Carrollian limit of the Milne worldsheet on its symmetry structure—analyzing the transformation properties of worldsheet fields in this limit may uncover novel aspects of tensionless string theory, including hidden symmetries or modified representations of worldsheet algebras.

Moreover, the emergence of time-like entanglement in this framework hints at potential connections between Carrollian physics and non-Lorentzian formulations of string theory. These ideas might find relevance in early universe scenarios and cosmological singularities, where conventional Lorentzian geometry may no longer provide an adequate description. Understanding how such structures integrate with broader string-theoretic models—including black hole interiors, Milne-type cosmologies, and anisotropic backgrounds—remains an exciting and largely unexplored frontier. While the current work lays the foundation, the following specific questions illustrate some promising directions:

\begin{itemize}

\item The Rindler worldsheet has already offered deep insight into near-horizon black hole physics, where accelerated observers experience phenomena such as the Unruh effect and Hawking radiation \cite{Socolovsky:2013rga}. In this context, the R and L Rindler wedges describe the exterior of black holes, while Carrollian and tensionless structures naturally emerge near the horizon \cite{Bagchi:2021ban,Bagchi:2023cfp,Bagchi:2024rje}. However, understanding the dynamics of the black hole interior remain elusive. Interestingly, the structure of the F and P Milne wedges—characterized by their time-like separation and strong anisotropic expansion—bears striking similarities to the near-singularity behavior inside black holes, where the classical Milne metric governs the interior evolution of spacetime \cite{Bueno:2024fzg}. Additionally, analytic solutions of certain black hole spacetimes indicate that their interiors develop Milne-like structures with possible time-like singularities \cite{Arean:2024pzo}. Furthermore, the Kruskal extension of black hole spacetimes suggests that the white hole region exhibits a structure closely related to the Milne sectors, reinforcing the importance of Milne physics not only for probing black hole interiors but also for understanding the evolution of white hole regions. This naturally raises several key questions: Can the null Milne worldsheet framework serve as an intrinsic probe of black hole interiors, just as the \textit{Rindler worldsheet in the Carrollian limit captures the exterior near-horizon physics? If so, what is the precise role of the characteristic time-evolution parameter in governing the non-inertial worldsheet dynamics inside a black hole?} Addressing these questions may offer a deeper examination of the Carrollian and tensionless limits of the Milne worldsheet and their possible implications for singularity resolution, black hole interiors, and non-Lorentzian gravity.

\item  Beyond black hole physics, the tensionless Milne worldsheet hints at a natural bridge to early universe cosmology. Near singularities, the early universe undergoes sequences of epochs, characterized by anisotropic expansion and contraction \cite{Belinsky:1970ew}, mirroring the structure of the Milne worldsheet. This suggests a deeper connection between non-inertial string dynamics and early universe physics. Moreover, Milne universes—describing expanding open cosmologies—share structural similarities with Kasner spacetimes in certain coordinate representations \cite{book:Birrell,mukhanov2005physical}, where their transformations reveal deep connections to non-inertial worldsheet formulations. The emergence of a Carrollian limit in the Milne worldsheet hints at new insights into ultra-relativistic regimes of early universe cosmology, particularly in settings where Carrollian symmetries may govern extreme spacetime dynamics. This raises fundamental questions: \textit{Can the tensionless limit of time-evolving strings provide insights into cosmic string behavior in an expanding Milne-like background? Could the interplay between non-inertial worldsheet evolution and Carrollian physics offer a novel perspective on early universe models and the quantum origins of large-scale structure?} Pursuing these ideas may illuminate the role of tensionless string dynamics in shaping the early universe and fundamental cosmological phenomena. 
\end{itemize}

Together, these questions suggest that the Milne framework might have broad relevance—from probing black hole interiors to exploring early-universe dynamics. While this paper lays the foundational groundwork, the avenues outlined above highlight how future work might further uncover the rich interplay between tensionless string theory, non-inertial evolution, and Carrollian physics.

\acknowledgments
The work of SK is supported by IIT Guwahati through Institute Post-Doctoral Fellowship.
	
\appendix
\section*{Appendix}
\section{Formulating the global modes for Milne worldsheet expansion} \label{FGM}
In this section, we outline the key steps to formulate the global (or Unruh-Minkowski) modes for the Milne worldsheet quantum mode expansion \eqref{KBT17}. We begin by setting up the right-moving sector of the positive-frequency global Milne modes. Utilizing the coordinate mapping \eqref{KME8}, we successively reexpress the local Milne modes $\tensor*[^{\rm F}]{\tilde{\mathcal{U}}}{_n}$, $\tensor*[^{\rm P}]{{\mathcal{U}}}{_n^*}$, $\tensor*[^{\rm P}]{\tilde{\mathcal{U}}}{_n}$, and $\tensor*[^{\rm F}]{{\mathcal{U}}}{_n^*}$ successively as follows:  
\begin{subequations}\label{KBT1}
\allowdisplaybreaks{\begin{align}
\sqrt{4\pi}n\,\tensor*[^{\rm F}]{\tilde{\mathcal{U}}}{_n} &= i\left(e^{\eta^-_{\rm F}a/c}\right)^{-i\Omega_{n}/a} = i e^{-i\Omega_{n}\psi}\left[a\left(\sigma^- - \varphi\right)/c\right]^{-i\Omega_{n}/a},\label{KBT1a}\\[6pt]
\sqrt{4\pi}n\,\tensor*[^{\rm P}]{{\mathcal{U}}}{_n^*}  &= -i\left(e^{-\eta^+_{\rm P}a/c}\right)^{-i\Omega_{n}/a} = -i\left(-1\right)^{-i\Omega_{n}/a} e^{-i\Omega_{n}\psi}\left[a\left(\sigma^- - \varphi\right)/c\right]^{-i\Omega_{n}/a},\label{KBT1b}\\[6pt]
\sqrt{4\pi}n\,\tensor*[^{\rm P}]{\tilde{\mathcal{U}}}{_n}  &= i\left(e^{-\eta^-_{\rm P}a/c}\right)^{i\Omega_{n}/a} = i e^{i\Omega_{n}\psi}\left[a\left(-\sigma^+ - \varphi\right)/c\right]^{i\Omega_{n}/a},\label{KBT1c}\\[6pt]
\sqrt{4\pi}n\,\tensor*[^{\rm F}]{{\mathcal{U}}}{_n^*} &= -i\left(e^{\eta^+_{\rm F}a/c}\right)^{i\Omega_{n}/a} = -i \left(-1\right)^{i\Omega_{n}/a}e^{i\Omega_{n}\psi}\left[a\left(-\sigma^+ - \varphi\right)/c\right]^{i\Omega_{n}/a}.\label{KBT1d}
\end{align}}
\end{subequations}
We then obtain the following mode combinations:
\begin{align}
\tensor*[^{\rm F}]{\tilde{\mathcal{U}}}{_n} - e^{-\pi \Omega_{n} /a}\,\tensor*[^{\rm P}]{{\mathcal{U}}}{_n^*} &= {i \over \sqrt{\pi}n}\left(a/c\right)^{-i\Omega_{n}/a}e^{-i\Omega_{n}\psi}\left(c\tau-\sigma - \varphi\right)^{-i\Omega_{n}/a},\label{KBT2}\\[6pt]
\tensor*[^{\rm P}]{\tilde{\mathcal{U}}}{_n} - e^{-\pi \Omega_{n} /a}\,\tensor*[^{\rm F}]{{\mathcal{U}}}{_n^*} &= {i \over \sqrt{\pi}n}\left(a/c\right)^{i\Omega_{n}/a}e^{i\Omega_{n}\psi}\left(-c\tau - \sigma - \varphi\right)^{i\Omega_{n}/a},\label{KBT3}
\end{align}
which are well-defined and analytically continued between the F and P wedges for the right-moving global Milne modes in Minkowski coordinates. Note that in steps \eqref{KBT1b} and \eqref{KBT1d} (corresponding to the negative frequency left-moving modes in P and F, respectively), there is an apparent ambiguity in the choice of $\left(-1\right)^{\mp i\Omega_{n}/a}: -1 = e^{i\pi}$ or $-1 = e^{-i\pi}$. However, it is crucial to note that $\sigma^{-} > 0$ ($< 0$) and  $\sigma^{+} > 0$ ($< 0$) in F (P). Thus, to maintain positive frequencies for the right-moving global mode combinations \eqref{KBT2} and \eqref{KBT3}, we must satisfy $\operatorname{Im}\,\sigma^- > 0$ in step \eqref{KBT1b} (continuing from P to F) and $\operatorname{Im}\,\sigma^- < 0$ in step \eqref{KBT1d} (continuing from F to P). In other words, the global mode \eqref{KBT2} is analytic in the upper half of the complex $\sigma^-$ plane, allowing us to set $-1 = e^{i\pi}$ in step \eqref{KBT1b}. Similarly, we set $-1 = e^{-i\pi}$ in step \eqref{KBT1d}, as the global mode \eqref{KBT3} is analytic in the lower-half complex $\sigma^-$ plane.
	
Next, we turn to the setup for left-moving sector of the positive-frequency global Milne modes. Using the light-cone coordinate mapping \eqref{KME8}, we can successively restructure the local Milne modes $\tensor*[^{\rm F}]{{\mathcal{U}}}{_n}$, $\tensor*[^{\rm P}]{\tilde{\mathcal{U}}}{_n^*}$, $\tensor*[^{\rm P}]{{\mathcal{U}}}{_n}$, and $\tensor*[^{\rm F}]{\tilde{\mathcal{U}}}{_n^*}$ as follows: 
\begin{subequations}\label{KBT4}
	\allowdisplaybreaks{\begin{align}
			\sqrt{4\pi}n\,\tensor*[^{\rm F}]{{\mathcal{U}}}{_n} &= i\left(e^{\eta^+_{\rm F}a/c}\right)^{-i\Omega_{n}/a} = i e^{-i\Omega_{n}\psi}\left[a\left(\sigma^+ + \varphi\right)/c\right]^{-i\Omega_{n}/a},\label{KBT4a}\\[6pt]
			\sqrt{4\pi}n\,\tensor*[^{\rm P}]{\tilde{\mathcal{U}}}{_n^*}  &= -i\left(e^{-\eta^-_{\rm P}a/c}\right)^{-i\Omega_{n}/a} = -i\left(-1\right)^{-i\Omega_{n}/a} e^{-i\Omega_{n}\psi}\left[a\left(\sigma^+ + \varphi\right)/c\right]^{-i\Omega_{n}/a},\label{KBT4b}\\[6pt]
			\sqrt{4\pi}n\,\tensor*[^{\rm P}]{{\mathcal{U}}}{_n}  &= i\left(e^{-\eta^+_{\rm P}a/c}\right)^{i\Omega_{n}/a} = i e^{i\Omega_{n}\psi}\left[a\left(-\sigma^- + \varphi\right)/c\right]^{i\Omega_{n}/a},\label{KBT4c}\\[6pt]
			\sqrt{4\pi}n\,\tensor*[^{\rm F}]{\tilde{\mathcal{U}}}{_n^*} &= -i\left(e^{\eta^+_{\rm F}a/c}\right)^{i\Omega_{n}/a} = -i \left(-1\right)^{i\Omega_{n}/a}e^{i\Omega_{n}\psi}\left[a\left(-\sigma^- + \varphi\right)/c\right]^{i\Omega_{n}/a}.\label{KBT4d}
	\end{align}}
\end{subequations}
In a similar manner, we derive the following mode combinations, analytically continued between the F and P wedges (i.e., along the $\sigma = 0$ surface) to produce the left-moving global Milne modes:
\begin{align}
		\tensor*[^{\rm F}]{{\mathcal{U}}}{_n} - e^{-\pi \Omega_{n} /a}\,\tensor*[^{\rm P}]{\tilde{\mathcal{U}}}{_n^*} &= {i \over \sqrt{\pi}n}\left(a/c\right)^{-i\Omega_{n}/a}e^{-i\Omega_{n}\psi}\left(c\tau + \sigma + \varphi\right)^{-i\Omega_{n}/a},\label{KBT5}\\[6pt]
		\tensor*[^{\rm P}]{{\mathcal{U}}}{_n} - e^{-\pi \Omega_{n} /a}\,\tensor*[^{\rm F}]{\tilde{\mathcal{U}}}{_n^*} &= {i \over \sqrt{\pi}n}\left(a/c\right)^{i\Omega_{n}/a}e^{i\Omega_{n}\psi}\left(-c\tau + \sigma + \varphi\right)^{i\Omega_{n}/a}.\label{KBT6}
\end{align}
As before, we encounter the ambiguity in choosing $\left(-1\right)^{\mp i\Omega_{n}/a}$ in steps \eqref{KBT4b} and \eqref{KBT4d}, which involve the negative frequency right-moving modes in the P and F regions, respectively. To ensure positive frequencies for the left-moving global mode combinations \eqref{KBT5} and \eqref{KBT6}, we require $\operatorname{Im},\sigma^+ > 0$ in step \eqref{KBT4b} (continuing from P to F) and $\operatorname{Im},\sigma^+ < 0$ in step \eqref{KBT4d} (continuing from F to P). Consequently, we set $-1 = e^{i\pi}$ in step \eqref{KBT4b} and $-1 = e^{-i\pi}$ in step \eqref{KBT4d}.

It is important to note that all the forms of the global Milne worldsheet modes derived in \cref{KBT2,KBT3,KBT5,KBT6} are unnormalized. To properly normalize these right- and left-moving global Milne modes, we need to adjust them to become orthonormal by determining the appropriate normalization constants. This process will yield the normalized forms of the global modes and their Hermitian conjugates for the Milne worldsheet, as presented in \cref{KBT16,KBT15}.
		
	
\bibliographystyle{JHEP}
\bibliography{biblio.bib}

\end{document}